# Quantum tunnelling-integrated optoplasmonic nanotrap enables conductance visualisation of individual proteins

**Authors:** Biao-Feng Zeng[1], Zian Wang[1], Yuxin Yang[1], Xufei Ma[1], Liang Xu[1], Yi Shen[1], Long Yi[1], Yizheng Fang[3], Ye Tian[4], Zhenrong Zheng[1], Yudong Cui[1], Ji Cao[3], Ge Bai[5], Weixiang Ye[6], Pan Wang[1], Cuifang Kuang[1], Joshua B. Edel[7], Aleksandar P. Ivanov[7], Xu Liu[1]*, Longhua Tang[1,2]*

**Affiliations:**

[1]State Key Laboratory of Extreme Photonics and Instrumentation, Interdisciplinary Centre for Quantum Information, College of Optical Science and Engineering, Zhejiang University, Hangzhou 310027, China.

[2]Nanhu Brain-Computer Interface Institute; Second Affiliated Hospital School of Medicine, Hangzhou, Zhejiang, 311100, China.

[3]Innovation Institute for Artificial Intelligence in Medicine, Zhejiang Province Key Laboratory of Anti-Cancer Drug Research, College of Pharmaceutical Sciences, Zhejiang University, Hangzhou, China.

[4]State Key Laboratory of Fluid Power and Mechatronic Systems, College of Mechanical Engineering, Zhejiang University, Hangzhou 310027, China.

[5]Department of Neurobiology and Department of Neurology of Second Affiliated Hospital, Zhejiang University School of Medicine, Hangzhou 310058.China.

[6]Center for Theoretical Physics, School of Physics and Optoelectronic Engineering, Hainan University, Haikou 570228, China.

[7]Department of Chemistry, Molecular Science Research Hub, Imperial College London, London, UK.

*Corresponding author. Email: liuxu@zju.edu.cn; lhtang@zju.edu.cn

**Abstract:** Biological electron transfer (ET) relies on quantum mechanical tunnelling through a dynamically folded protein. Yet, the spatiotemporal coupling between structural fluctuations and electron flux remains poorly understood, largely due to limitations in existing experimental techniques, such as ensemble averaging and non-physiological operating conditions. Here, we introduce a quantum tunnelling-integrated optoplasmonic nanotrap (QTOP-trap), an optoelectronic platform that combines plasmonic optical trapping with real-time quantum tunnelling measurements. This label-free approach enables single-molecule resolution of protein conductance in physiological electrolytes, achieving sub-3 nm spatial precision and 10-μs temporal resolution. By synchronising optoelectronic measurements, QTOP-trap resolves protein-specific conductance signatures and directly correlates tertiary structure dynamics with conductance using a "protein switch" strategy. This methodology establishes a universal framework for dissecting non-equilibrium ET mechanisms in individual conformational-active proteins, with broad implications for bioenergetics research and biomimetic quantum device design.

## Introduction

Biological electron transfer (ET) processes[1-4] drive essential biological phenomena, from photosynthesis to oxidative phosphorylation, via quantum tunnelling mechanisms across protein domains[5-8]. These mechanisms are evolutionarily optimised to maximise electron flux while encoding functional states through coupled electronic networks. Deciphering the relationships between structure and conductance at the single-protein level is a fundamental biophysical challenge, with significant implications for both understanding energy transduction and advancing quantum bioelectronics[9].

Although techniques such as cryo-EM[10], nanopore sensing[11-13], evanescent scattering microscopy[14, 15], and fluorescence resonance energy transfer (FRET)[16-18] have transformed protein structural biology, they are insufficient for probing quantum aspects of ET. Single-molecule spectroscopy can detect ET-related spectral shifts[7, 19], however lacks the temporal resolution or physiological compatibility needed to resolve native tunnelling events. Proteotronics aims to bridge this gap by integrating proteins into electronic circuits, however, early approaches using electrode-immobilised proteins suffered from ensemble averaging, while single-molecule scanning tunnelling microscopy[20, 21] and nanogap electronics[22-25] require covalent anchoring or engineered binding sites, compromising protein conformational integrity. These limitations highlight the need for a label-free, high-throughput platform capable of monitoring protein conductance in native physiological conditions.

## Results

### *QTOP-trap Platform Design and Validation.*

We developed QTOP-trap by integrating plasmonic optical trapping with real-time tunnelling spectroscopy, enabling high-resolution conductance profiling of individual proteins in solution (Fig. 1A). The platform employed quantum mechanical tunnelling (QMT) probes, containing gold nanoelectrode pairs fabricated via feedback-controlled electrodeposition on double-barrel nanopipettes (tip diameter ≈ 200 nm; Fig. 1B; fig. S1)[26, 27]. Current-voltage characterisation in PBS (−0.3 to +0.3 V; Fig. 1C; fig. S2) revealed nonlinear tunnelling behaviour consistent with quantum transport, with gap distances of 0.5 - 3 nm as estimated by the Simmons model[28]. Due to entropic and diffusional limitations, passive capture of single molecules in nanoscale gaps is inefficient, particularly at low analyte concentrations[29, 30], necessitating active transport[31-33]. To address this, we incorporated plasmonic optical trapping. Dark-field spectroscopy of the nanogap confirmed broadband plasmon resonances (583-751 nm; Fig. 1D). Finite-difference time-domain (FDTD) simulations predicted >$10^6$ fold electromagnetic field enhancement at 637 nm excitation (1 mW/μm², 0.5 nm gap), generating optical potential wells (1.1 $kT_0$ at gap centre; Fig. 1E, F) exceeding thermal fluctuations (~ $kT_0$)[34, 35]. Vector field analysis revealed directional gradient forces ($F_X$ ≈ 11 pN), capable of guiding polar molecules such as 4-pentyl-4'-cyanobiphenyl (5CB) toward the nanogap.

**Figure 1**

To differentiate tunnelling currents from plasmon-induced photocurrents, the system integrates patch-clamp detection with lock-in amplification (Fig. 2A; fig. S8). Polarised 637 nm illumination induced picoampere-scale AC photocurrents (Fig. 2B, D; fig. S9, S10), confirming effective plasmon–electrode coupling[36]. The laser-induced thermal heating was not considered, as the measurements were performed in solution and the macroscopic unilluminated portions of the electrodes could serve as heat sinks[37, 38]. Importantly, AC interference remained below 1% of the DC tunnelling signal at 200 mV bias, ensuring high signal-to-noise fidelity even under prolonged illumination in aqueous environments. Three-dimensional photocurrent mapping was employed to optimise optical alignment (Fig. 2C, fig. S11).

The QTOP-trap exhibited size-independent trapping capabilities across a diverse range of molecules (Fig. 2E, F; fig. S12-S14). For example, 5CB (length: ~1.7 nm), $[Ru(bpy)_3]^{2+}$ (principal axis: ~2.0-2.5 nm), and $C_{60}$ (diameter: ~0.71 nm) exhibited trapping behaviours correlated with their molecular polarisabilities. Active plasmonic trapping enhanced single-molecule detection probabilities by 43-fold (5CB, 1 μM), 16-fold ($[Ru(bpy)_3]^{2+}$, 1nM), and 13-fold ($C_{60}$, 5 μM) compared to passive capture (fig. S12C-E). In particular, 5CB's conductance distribution narrowed with increasing optical intensity, attributed to plasmon-induced molecular dipole alignment (fig. S15-S17). Furthermore, 10 nm gold nanoparticles, substantially larger than the tunnelling gap, displayed sustained current increase under plasmonic illumination (fig. S12; fig. S18).

**Figure 2**

### *Single-Protein Conductance Profiling in Physiological Solution.*

To test QTOP-trap's capability for native protein analysis, we repeatedly formed single-protein junctions in solution. This is a significant advantage compared to STM, which often relies on engineered surface anchors for molecular attachment[22, 39]. Instead, we introduced cysteine-functionalized nanoelectrodes to minimise non-specific gold-protein interactions and enhance conductance signals. This approach relies on hydrogen bonding between cysteine and surface-exposed polar moieties (e.g., hydroxyl or amine groups) on the protein exterior (Fig. 3A)[40]. Conductance-time traces of catalase (20 μg/mL in 1 mM PBS, pH 7.4) revealed two distinct regimes (Fig. 3B). Under dark conditions, baseline current fluctuations were accompanied by only sporadic, low-amplitude peaks - indicative of transient cysteine-protein interactions. Upon illumination with 5 mW laser power, prolonged high-current plateaus emerged ($\tau = 1.94 \pm 0.02$ ms, N = 12467 events), consistent with plasmon-induced optical trapping, which increases the frequency and dwell time of molecular binding events.

The complex features observed in these conductance traces are attributed to multivalent interactions between protein residues and the nanoelectrode surface. To quantify individual molecular events, discrete current spikes were extracted, with the temporal midpoint of each peak defined as $t_0$ (Fig. 3C). The dwell-time histograms followed a single-exponential decay distribution (fig. S19), characteristic of single-molecule dynamics[41]. Persistent high-current states under elevated irradiation were classified as stable trapping events. For each, the conductance was calculated as $\Delta G = \Delta I/V_{bias}$, where $V_{bias}$ ranged from 0.1 to 0.2 V, and normalised against the pre-event baseline current ($I_0$). Across $>10^3$ events from five native proteins (catalase, horseradish peroxidase, glucose oxidase, calmodulin, and S-protein), we constructed two-dimensional conductance-frequency histograms that revealed protein-specific signatures with well-defined Gaussian distributions (Fig. 3D, E; fig. S20; table S1). These clustering patterns enabled robust molecular discrimination based on conductance fingerprints.

To address device-to-device variability, we recalibrated baseline currents between different protein types, adopting a single-device, sequential measurement protocol. We achieved reproducibility within 95% across more than 10 device replicates, enabling accurate comparisons of conformational or mutational differences. As a proof-of-principle, we analysed wild-type glycyl-tRNA synthetase (GlyRS$^{WT}$) and its P234KY mutant (GlyRS$^{P234KY}$), which introduces a surface-exposed lysine substitution[42] (fig. S21, S22). In single-component solutions, GlyRS$^{P234KY}$ exhibited a ~2.3-fold higher median conductance (25.4 nS vs. 11.1 nS for GlyRS$^{WT}$), likely due to enhanced charge delocalisation and improved electron tunnelling via the lysine side chain. Binary mixture analysis revealed a clear bimodal distribution, confirming the QTOP-trap's ability to resolve single-protein conductance with high specificity under physiological, label-free conditions.

**Figure 3**

### *Dynamic Conformational Tracking with "Protein Switch"*

Although single proteins can be transiently trapped between nanoelectrodes, continuous monitoring of their conductance is often hindered by heterogeneous molecular orientations and short-lived binding events, which limit reliable tracking of conformational dynamics[43, 44]. To overcome this limitation, we introduce a "protein switch" architecture in which proteins are unidirectionally tethered to nanoelectrodes. This configuration enables continuous observation of single-protein conductance over extended periods (Fig. 4A).

As a proof of concept, we employed histidine-tag (His-tag) chemistry, an extensively validated immobilisation strategy, to achieve directional protein attachment via $Cu^{2+}$-NTA coordination[45]. Sequential functionalization of gold electrodes - NTA monolayer assembly, $Cu^{2+}$ chelation, and His-tagged protein binding - establishes a "molecular switch" arrangement in which the protein remains flexibly tethered at one terminus, while the remainder of the molecule is free to explore conformational space relative to the electrode surface. Fig. 4B illustrates the optically modulated trapping dynamics for WD repeat domain 5 protein (WDR5). Upon laser illumination, baseline current signals with sparse spikes transitioned into sustained current increases (fig. S23), indicating that optical forces guided the free protein terminus into the nanogap to form a stable junction. Deactivation of the laser restored the baseline current via Brownian-mediated dissociation, confirming the system's stability and reversibility (Fig. 4C; fig. S24). Treatment with EGTA, a calcium chelator that also sequesters $Cu^{2+}$, abolished His-tag-associated signals ($\Delta I/I_0 < 1\%$), confirming that coordination chemistry governs interfacial stability (fig. S25B).

To evaluate linker-specific effects, we engineered asymmetrically tagged WDR5 constructs with either His-tag or mono-streptavidin (MSA) at the free terminus (fig. S25C; fig. S26). Parallel experiments revealed similar conductance distributions (His-tag: 0.30 ± 0.05 μS; MSA-tag: 0.21 ± 0.02 μS), indicating minimal linker influence on electronic transport in the switch configuration (fig. S27). However, biotin-MSA-tethered proteins exhibited a delayed return to baseline conductance following laser deactivation (fig. S28), attributed to MSA's rigid structure hindering conformational relaxation.

We next applied this controlled-molecule switching strategy to monitor real-time conformational dynamics of the heat shock protein Hsp90. Hsp90 is a cytoplasmic chaperone that undergoes ATP-driven conformational changes between open and closed states, a critical component of its regulatory function in proteostasis. His-tagged Hsp90 was immobilised on $Cu^{2+}$-NTA-modified electrodes[46, 47]. In the absence of light, tunnelling currents showed only minor fluctuations, consistent with sporadic single-end collisions (fig. S29A). Upon 7 mW laser excitation, stable junctions formed rapidly, followed by sustained current oscillations (Fig. 4D-i, fig. S29B) attributable to internal conformational rearrangements of the anchored protein. The addition of ATP to pre-formed junctions enhanced current fluctuation frequency (Fig. 4D-ii; fig. S29C) and induced discrete bimodal conductance states (Fig. 4E) consistent with conformational switching between functional states. In contrast, the non-hydrolyzable ATP analogue AMP-PNP failed to induce such fluctuations (Fig. 4D-iii; fig. S29D), while EDTA-mediated $Cu^{2+}$ chelation disrupted His-tag anchoring and reset the current to baseline (fig. S29E, F).

Statistical analysis revealed two distinct conductance populations: state 1 at 423.86 ± 0.37 nS and state 2 at 514.74 ± 0.13 nS, corresponding to the open and closed conformations, respectively. Extracted transition rates ($k_{1\rightarrow 2} = 0.88 \pm 0.02$ ms$^{-1}$ and $k_{2\rightarrow 1} = 2.26 \pm 0.06$ ms$^{-1}$) were consistent with those reported for the native ATPase cycle of Hsp90[48, 49] (Fig. 4F, G). These data demonstrate the platform's capability to resolve single-protein conformational transitions and their associated electronic signatures in real-time.

**Figure 4**

## Discussion

In summary, we introduce the QTOP-trap platform—an optoelectronic system that enables simultaneous manipulation and label-free conductance analysis of individual proteins in physiological aqueous environments. By combining plasmonic optical trapping with nanoscale quantum tunnelling detection, QTOP-trap provides a dual-function strategy that allows both spatial confinement and electrical interrogation of single biomolecules. The "protein switch" extension further enhances this capability by stabilising protein-electrode interfaces through directional anchoring, enabling dynamic conformational tracking with molecular specificity. This strategy allows a direct correlation between protein structural transitions and their electronic transport properties. Importantly, the system supports high-resolution measurements without the need for chemical labelling or cryogenic conditions. The platform's ability to resolve dynamic protein states at the single-molecule level may open new avenues for studying protein folding landscapes, engineering biohybrid quantum systems, and elucidating the molecular basis of electron transfer in biological systems.

### Supplementary data:

Supplementary data are available at NSR online

### Funding:

This work was supported by the National Natural Science Foundation of China (grant no. 62127818, 22374129 and 62125504), the Natural Science Foundation of Zhejiang Province (grant no. LR22F050003), the Leading Innovative and Entrepreneur Team Introduction Program of Zhejiang (grant no. 2024R0100)

**Author contributions:** L.T., X.L., and B.-F.Z. conceptualised the experiments. B.-F.Z., Z.W. Y.S. performed the single-molecule experiments. B.-F.Z. and L.X. constructed the smPNT setup. B.-F.Z., Z.W, X.M., Y.T., and L.Y. prepared the QMT probes. P.W. and W.Y. performed the dark-field imaging. Y.Y. and B.-F.Z conducted the data analysis. Y.F. synthesised the WDR5 proteins. G.B. provided the GlyRS proteins. B.-F.Z carried out the FDTD simulation, with help from X. M. performed the calculation of molecular polarizability. L.T., X.L., and B.-F.Z wrote the manuscript. L.T., X.L., J.B.E, A.P. I, C.K., P.W., W.Y., J.C., G.B., Z.Z and Y.C. reviewed and edited the manuscript.

**Conflict of interest statement. None declared.**

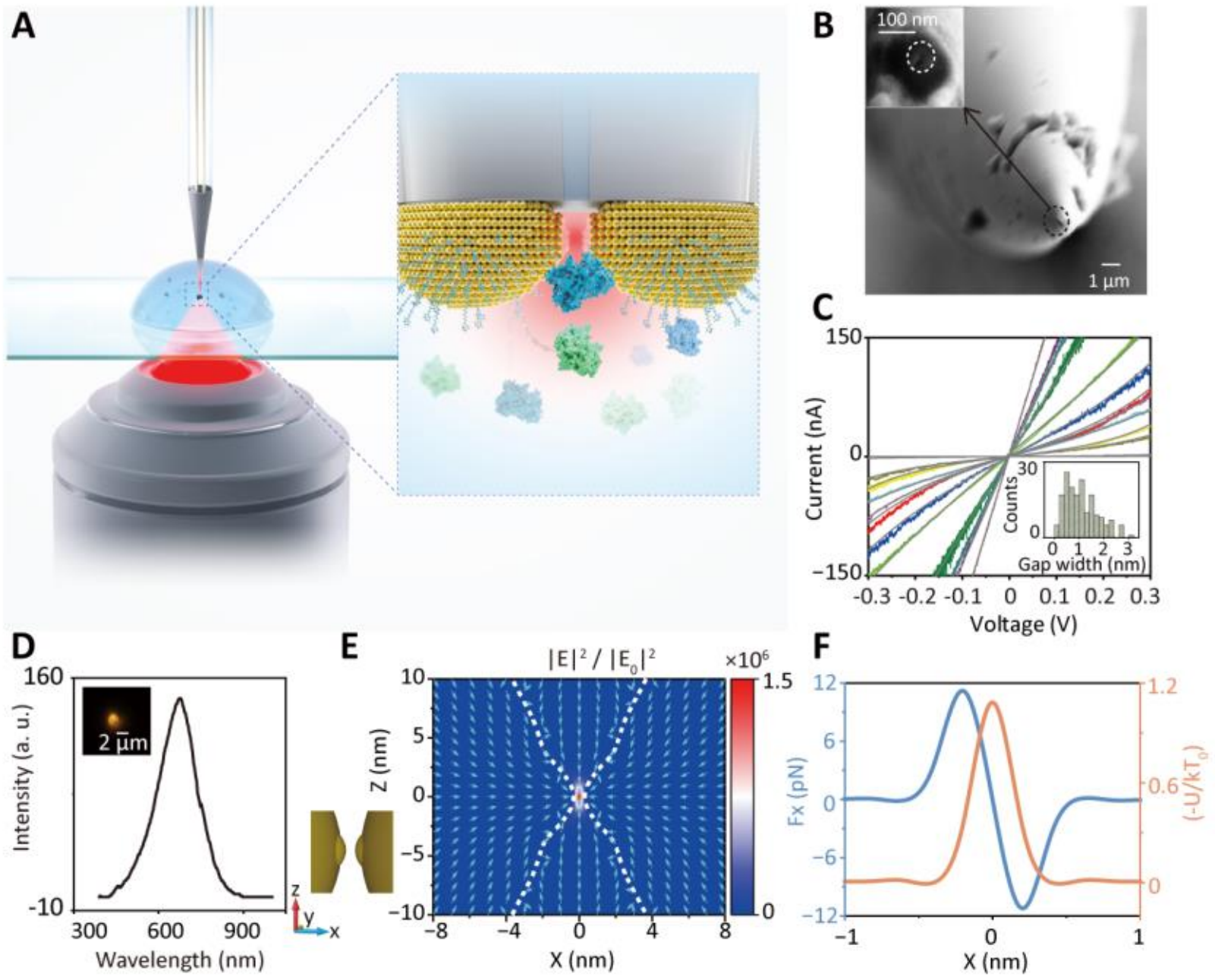

**Fig. 1. Design and theoretical basis of quantum tunnelling-integrated optoplasmonic nanotrap (QTOP-trap). (A)** Schematic illustration of the QTOP-trap platform. A quantum mechanical tunnelling (QMT) probe is immersed in solution and illuminated with focused light to induce a plasmonic "hot spot." Target molecules are optically trapped within the nanogap and detected through tunnelling current fluctuations. **(B)** Scanning electron microscopy (SEM) images of the QMT probe reveal the nanoscale conical tip geometry. The tunnelling gap itself remains below SEM resolution. **(C)** Tunnelling current-voltage (I-V) plots of QMT probes in deionised (DI) water at room temperature. Coloured dots represent experimental measurements; grey curves show Simmons model fits. Inset: histogram of estimated gap distances for 170 QMT probes, derived from Simmons model fitting. **(D)** Dark-field scattering spectra show a resonance peak ranging from 583 nm to 751 nm (FWHM). Inset: light scattering at the gold-coated tip, indicating plasmonic activity. **(E)** Spatial distribution of the calculated electric field enhancement factor, **(F)** x component of the transverse optical force $F$x (blue curve) and trapping potential cross-section (orange curve) exerted on the 5CB molecule at the X-Z plane (gap distance: 0.5 nm) upon 637 nm illumination wavelength with an intensity of 1 mW/μm² (see figs. S3-S7 for detailed simulation results). The blue arrows in **(E)** show the mapping of the normalised optical force vector, indicating that the target molecules near the nanoscale gap will move towards the local hotspots under the action of the optical gradient force. The grid of the optical force vector is processed using a sparse method, with a sparse step of 5. The trapping potential U is expressed in thermal energy units $kT_0$ ($T_0$ = 300 K).

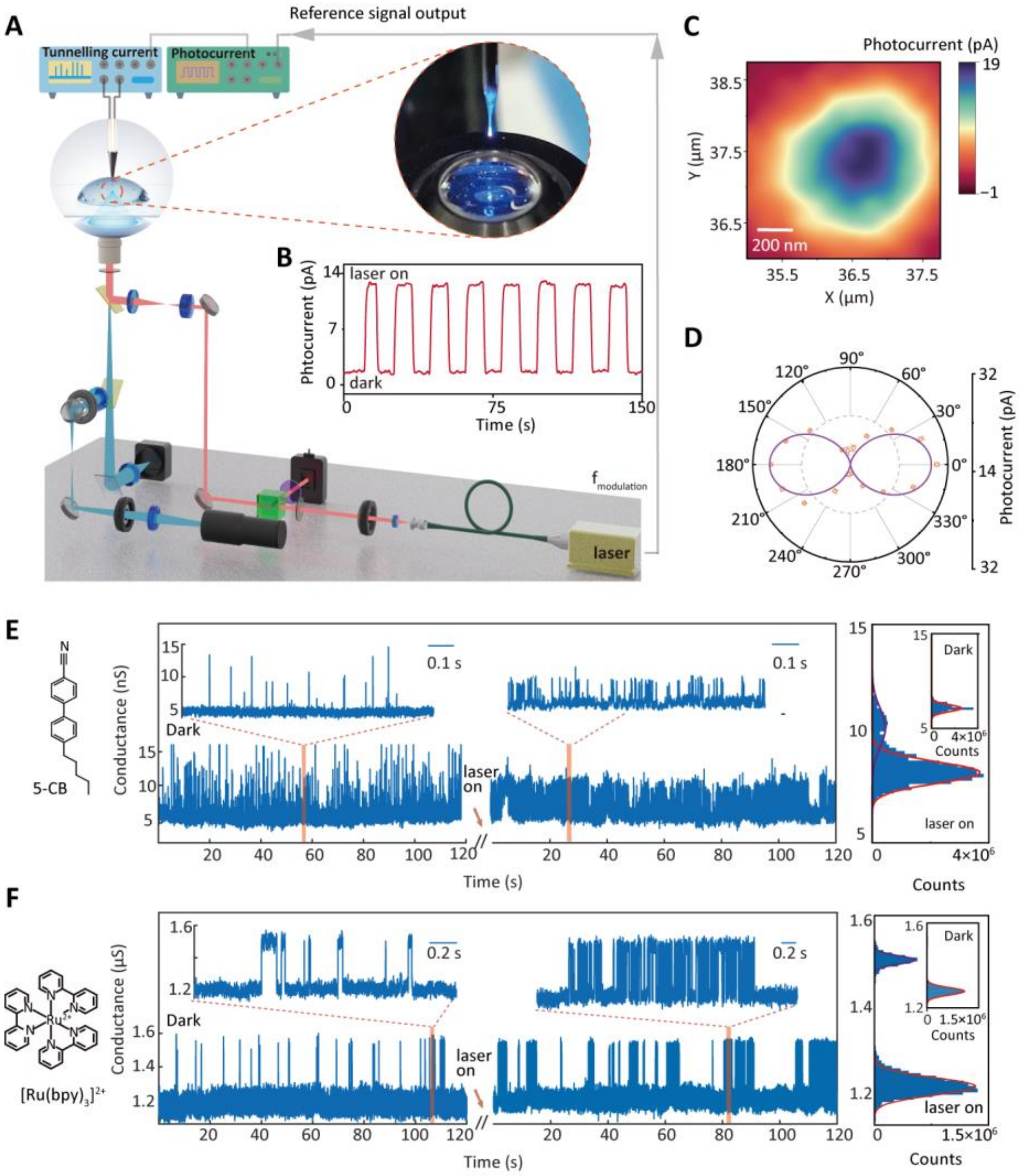

**Fig. 2. Optoelectronic measurement with QTOP-trap platform. (A)** Schematic of the plasmonic optical trapping combined with real-time quantum tunnelling measurements. A polarised 637 nm laser beam was collimated and focused on the tip of the QMT probe. For photocurrent detection, the light is amplitude-modulated by an arbitrary waveform generator (AWG), with the modulation frequency serving as a reference for lock-in detection. For tunnelling current measurements, the laser operates in continuous wave (CW) mode to excite localised surface plasmon resonances, facilitating optical trapping and enabling tunnelling detection via a patch-clamp amplifier. Inset: bright-field image of the illuminated QMT probe. **(B)** Photocurrent trace showing signal modulation in response to laser on/off switching using a mechanical shutter, confirming stable plasmonic excitation. Parameters: 1077 Hz modulation frequency, 2 mW laser power, 0 mV bias, in DI water. **(C)** Two-dimensional photocurrent map acquired by raster-scanning the QMT probe across the X–Y plane while maintaining a stationary laser focus. The probe was mounted on a 3D nanopositioning stage. Scan step: 250 nm; modulation frequency: 1077 Hz; laser power: 1.1 mW; bias: 0 V, in air. **(D)** Photocurrent amplitude as a function of polarisation angle, fitted with a $\cos^2\theta$ fitting (solid line). The polarisation angle of 0° was adjusted to the position where the photocurrent reached a maximum, indicating that the polarisation was aligned across the nanogap. Modulation frequency: 1077 Hz, laser power: 2 mW, bias: 0 V, in DI

water. **(E-F)** Representative conductance traces and histograms of **(E)** 5CB at 200 mV and **(F)** $[Ru(bpy)_3]^{2+}$ at 100 mV, recorded with laser-off and laser-on, reveal an increased number of single-molecule events after the light activation. The molecular structures of 5CB and $[Ru(bpy)_3]^{2+}$ are shown in the left panel, and the inset zoom-in traces display the clear transient conductance jumps, indicating the presence of a single molecule within the tunnelling junction.

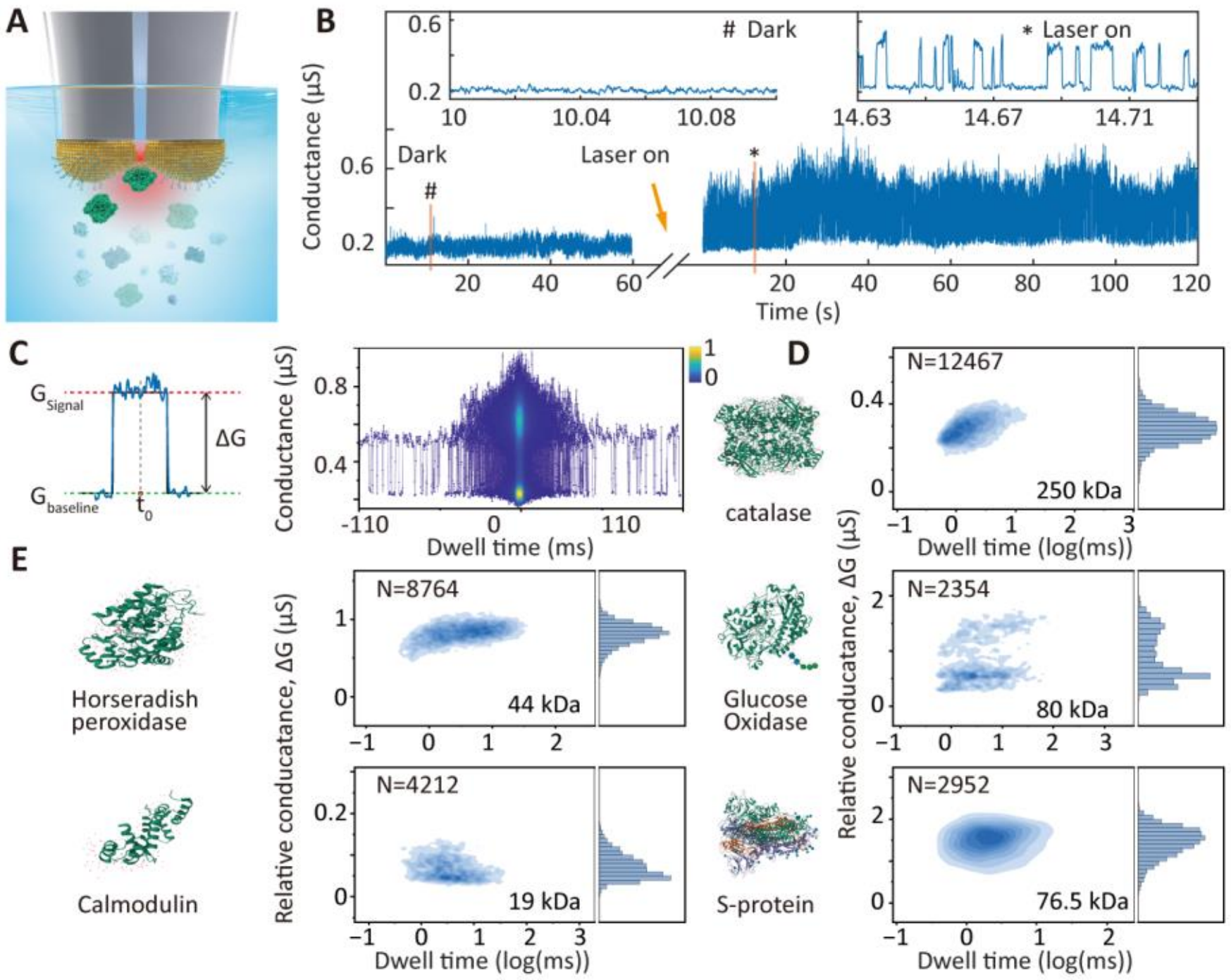

**Fig. 3. Conductance profiling of single proteins in aqueous solution. (A)** Schematic illustration depicting the capture of freely diffusing protein molecules into the tunnelling nanogap by optical trapping for subsequent conductance measurements in solution. **(B)** Representative conductance traces of catalase acquired at 100 mV bias under dark and illuminated conditions (laser power: 5 mW; buffer: 1 mM PBS, pH 7.4). Insets: expanded 0.1s segments of the traces showing discrete current fluctuations. **(C)** Representative peak signal extracted from the conductance trace in **(B)** and the 2D density map of peak conductance signals extracted from all experimental traces under illumination. The zero point on the x-axis denotes the time position in the middle of each peak signal, as marked as $t_0$. **(D)** 2D density scatter plot of relative conductance (ΔG) versus the logarithm of dwell time for all captured events. ΔG is computed as the difference between the peak and baseline current, normalised by the applied voltage (as defined in panel (**C**)). Left panel: crystal structure of catalase (sourced from the Protein Data Bank). Right panel: histogram showing the distribution of ΔG values. **(E)** Comparative conductance profiling of four additional proteins: horseradish peroxidase, glucose oxidase, calmodulin, and S-protein, recorded under laser illumination.

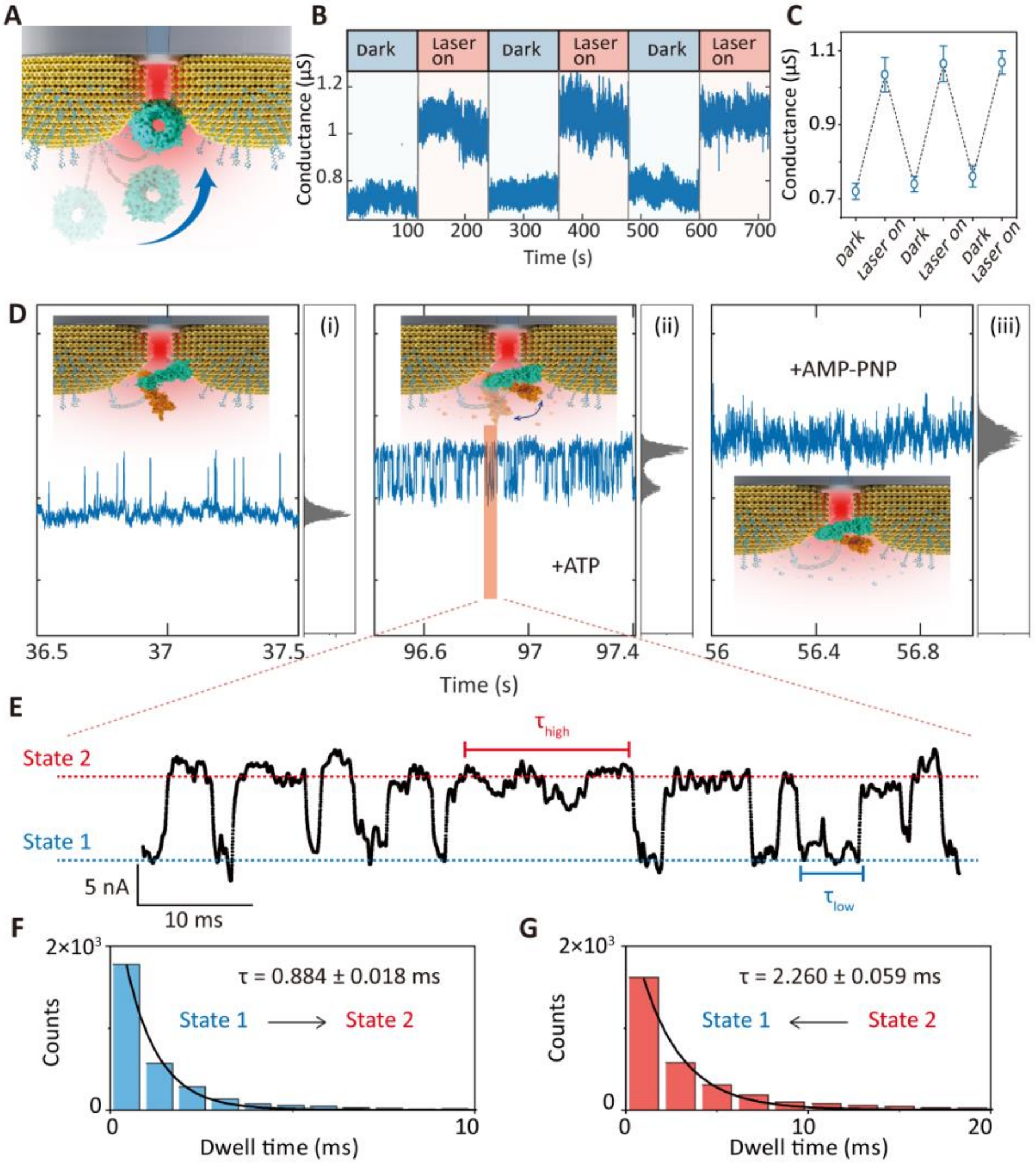

**Fig. 4. Tracking protein conformational dynamics with "protein switch" strategy. (A)** Schematic representation of the single-molecule "protein switch" architecture. Proteins were unidirectionally tethered via site-specific functionalization to one end of the QMT probe and measured in 1 mM PBS (pH 7.4). In the absence of illumination (laser off), the tethered protein freely swung in solution. Upon laser activation (laser-on), optical gradient forces drove the protein into the tunnelling nanogap, establishing a reversible single-molecule junction. Repeated laser modulation enabled dynamic capture and release, mimicking switch-like motion. **(B)** Trapping and release of His-tag-linked WDR5 protein junctions were performed by alternatively turning the light on and off, demonstrating the reversibility of protein junctions. **(C)** Plot of conductance values extracted from the conductance traces of **(B)**. Each point represents the mean conductance of traces, with error bars representing the ±SD from the mean. **(D)** Representative conductance traces of Hsp90 protein junctions recorded under various conditions: (i) laser-on, (ii) ATP addition, and (iii) AMP-PNP (non-hydrolysable ATP analogue). Measurements were conducted at 100 mV bias, with 7 mW laser power in 1 mM PBS (pH 7.4). Inset: illustration of Hsp90 conformational changes between open and closed states upon the addition of two types of nucleotides. **(E)** Representative conductance trace following ATP addition. Two discrete conductance levels of Hsp90 are indicated: low (blue dashed line, open state) and high (orange dashed line, closed state). Dwell times representative conductance trace with ATP. The low conductance state (blue dashed line) and high conductance state (orange dashed line) represent the open and closed states. Dwell

time in each state ($\tau_{high}$ and $\tau_{low}$) was extracted using K-means clustering of peak events. **(F-G)** Histograms of conductance values corresponding to **(F)** the low (open) and **(G)** high (closed) conformational states of Hsp90 under ATP conditions. The transition kinetics were determined by a single-exponential fit of the dwell time distributions.

Supplementary Materials for

# Quantum tunnelling-integrated optoplasmonic nanotrap enables conductance visualisation of individual proteins

Biao-Feng Zeng[1], Zian Wang[1], Yuxin Yang[1], Xufei Ma[1], Liang Xu[1], Yi Shen[1], Long Yi[1], Yizheng Fang[3], Ye Tian[4], Zhenrong Zheng[1], Yudong Cui[1], Ji Cao[3], Ge Bai[5], Weixiang Ye[6], Pan Wang[1], Cuifang Kuang[1], Joshua B. Edel[7], Aleksandar P. Ivanov[7], Xu Liu[1]*, Longhua Tang[1,2]*

Corresponding author: liuxu@zju.edu.cn; lhtang@zju.edu.cn

**The PDF file includes:**

Materials and Methods
Supplementary Text
Figs. S1 to S31
Tables S1
References

## Materials and Methods

Materials

The L-cysteine, Tris(2-carboxyethyl) phosphine hydrochloride (TCEP), Dimethyl sulfoxide (DMSO), Calmodulin (CaM), Adenosine 5'-triphosphate (ATP) disodium salt hydrate, Adenosine 5′-(β,γ-imido)triphosphate lithium salt hydrate (AMP-PNP) were obtained from Sigma-Aldrich (Shanghai) Trading Co., LTD. Biotin PEG thiol (biotin), Ethylenediaminetetraacetic acid (EDTA), Glucose Oxidase (GOD), and Catalase (CAT) were purchased from Shanghai Aladdin Biochemical Technology Co., LTD. Horseradish Peroxidase (HRP) was purchased from Shanghai solarbio Bioscience & Technology Co., LTD. SARS-CoV-2 (2019-nCoV) Spike S1-His Recombinant Protein (S-protein) was purchased from Sino Biological (China). Heat shock protein 90 alpha family class A member 1 (Hsp90) was purchased from Sangon Biotech (Shanghai) Co., Ltd. WD repeat domain 5 protein (WDR5) was purified in the lab. The wild-type glycyl-tRNA synthetase ($GlyRS^{WT}$) and the missense mutation glycyl-tRNA synthetase ($GlyRSP^{234KY}$) were prepared as previously described[1]. Tris(2,2-bipyridyl)ruthenium(II) chloride hexahydrate ($[Ru(bpy)_3]^{2+}$) was purchased from Shanghai Xianding Biotechnology Co., LTD. Au nanoparticles (AuNPs, 10 nm) were purchased from Nanjing XFNANO Materials Tech Co., LTD. Copper(II) sulfate pentahydrate ($CuSO_4 \cdot 5H_2O$) was purchased from Sinopharm Chemical Reagent Co., LTD. Dithiobis (C2-NTA) was purchased from DOJINDO Laboratories. 4-pentyl-4'-cyanobiphenyl (5CB) was purchased from Bide Pharmatech Ltd. Fullerene-$C_{60}$ ($C_{60}$) was purchased from Shanghai Macklin Biochemical Technology Co., LTD. Phosphate buffer solution (PBS, 10 mM, pH 7.2-7.4, without $Ca^{2+}$/$Mg^{2+}$) was purchased from SinoDetech Scientific, LTD. The plating solution was commercially purchased from Tianyue New Material (Au concentration 1-3 g/L). All other chemicals used in the fabrication of quantum mechanical tunnelling (QMT) probes were sourced from Sigma-Aldrich. Deionised (DI) water with a resistivity of 18.2 MΩ cm was used in all the experiments.

Fabrication of the quantum mechanical tunnelling probes

Quantum mechanical tunnelling (QMT) probes were fabricated using double-barrel quartz capillaries (outer diameter: 1.2 mm; inner diameter: 0.90 mm; length: 100 mm), following established procedures. A laser puller (P-2000, Sutter Instrument) was employed to locally heat the capillary centre while applying axial tension, yielding two tapered probes with tip diameters of ~100 nm. Carbon pyrolytic decomposition was subsequently used to deposit conductive carbon within the barrels (~1-2 mm from the tip), forming internal electrodes connected to external copper leads. Electrochemical etching was then carried out to selectively remove carbon at the tips, electrically isolating the two barrels. Finally, feedback-controlled electrochemical deposition was implemented to create a sub-5-nm tunnelling gap between the electrodes. This process involved real-time current monitoring between the nanoscale electrodes. Deposition was automatically stopped when the current exceeded 10 pA under a 100 mV bias. Completed QMT probes were stored in deionised (DI) water to facilitate self-reorganisation and stabilisation of the surface gold atoms. Gap dimensions were characterised via current-voltage measurements and fitted using the Simmons tunnelling model.

Experimental setup

The optical setup comprises three optical path assemblies: excitation, illumination, and imaging (Supplementary Fig. 9). Briefly, a laser diode (Coherent OBIS LX637), serving as the light source, provides the incident wavelength of 637 nm. For photocurrent measurement, the laser diode was driven by a square-wave signal to generate an intensity-modulated excitation light. Light from the laser diode was collimated and conditioned by the optical elements, including irises, achromatic lenses, reflectors, beamsplitters, a half-wave plate, and a 40× objective (LUCPlanFLN, Olympus, NA=0.6), then focused on the tip of the QMT probes. The half-wave plate, positioned anterior to the objective lens, minimises polarisation axis perturbations from optical components, preserving linear polarisation integrity. The illumination subsystem employed the Köhler configuration, where light from a light-emitting diode (LED) was conjugated to the entrance pupil plane of the objective, thereby generating homogeneous illumination. The reflection light was collected by an area-array camera (MV-CA050-20UC, Hikrobot), facilitating the detection of the reflection spot from the tip of the QMT probes.

Photocurrent signals were acquired using a lock-in amplifier (MFLI, Zurich Instruments), with the frequency of the driven laser diode signal functioning as the reference frequency for the photocurrent detection demodulated with a 100 ms time constant, achieving an optimised signal-to-noise ratio (SNR). In the photocurrent scanning experiment, the QMT probe was mounted on a parallel-kinematic 3-dimensional cube nanopositioner (P-616.3, Physik Instrumente). The laser was initially aligned to the tip of the QMT probe using the reflection spot. Subsequently, the position of the QMT probe was manually fine-tuned based on the feedback from the photocurrent signal measured by the lock-in amplifier. Finally, spatial scanning was performed by controlling the position of the QMT probe via the nanopositioner to obtain the spatial distribution map of the photocurrent. To enhance the SNR of the photocurrent measurement, the integration time of the lock-in detection during the scan was set to 100 ms with a sampling rate of 1.673 kHz. Each scanning point was sampled for 2 seconds, and the average value over the entire sampling duration was used as the signal for that point. For tunnelling current detection, the laser diode was driven by direct current (DC) to maintain the stability of plasmonic field excitation. The tunnelling current signal was obtained by utilising MultiClamp 2400 (Molecular Devices) operated in voltage-clamp mode and digitised using Axon Digidata 1550B with a bandwidth of 100 kHz.

Flow cell preparation

A quartz capillary tube with an inner diameter of 1.5 mm served as a flow cell. Specifically, the quartz capillary and the QMT probe were coaxially and vertically mounted in a custom-designed holder. The QMT probe extended into the interior of the capillary tube and was positioned roughly 1 mm above the bottom. The bottom of the capillary tube contacted a hydrophobic cover glass, which acted as the substrate. During the measurement, the target solution was introduced through the inlet at the top of the capillary tube and flowed downward along the inner wall under gravity, submerging the tip of the QMT probe. If the solution continues to be injected, it will eventually flow out from the bottom of the capillary. In this case, the excess solution at the bottom could be removed using a micropipette, enabling the replacement of the target solution or rinsing the QMT probe.

Single-molecule conductance measurements via free collision strategy.

For molecule and nanoparticle detection, $[Ru(bpy)_3]^{2+}$, 5CB, $C_{60}$, and AuNPs were sensed on bare tunnelling electrodes. $[Ru(bpy)_3]^{2+}$ (1 nM) and AuNPs (500 ng/mL) were dissolved in DI water, 5CB (1 μM) was dissolved in a mixed solvent of ethanol and water (1:1, v/v), and $C_{60}$ (3.6 μg/mL, 5 μM) was dissolved in DMSO. In the experiments, 40 μL of the prepared target solution was dropped into the inlet of the flow cell, submerging the tip of the tunnelling electrodes. For conductance measurement of proteins in aqueous solution, cysteine-functionalized tunnelling electrodes were employed to assay proteins. The QMT probe was initially cleaned in DI water and ethanol, then incubated in 0.2 mM cysteine for over 4 hours to functionalize the surface of tunnelling electrodes with cysteine. After processing, the tunnelling probe was rinsed with DI water to remove the non-covalently bound cysteine. Subsequently, the cysteine-functionalized tunnelling electrodes were used for conductance measurements of free proteins in solution. CaM (1 μg/mL), S-Protein (1 μg/mL), HRP (2 μg/mL), GOD (3.5 μg/mL), and CAT (20 μg/mL) were diluted in 1 mM PBS (pH 7.4), and subsequently introduced into the flow cell for conductance measurements. For single-molecule plasmonic optical trapping experiments, the modulated excitation light was focused on the apex of the tunnelling electrode. Upon excitation, the tunnelling junction generates a photocurrent, which is fed into a lock-in amplifier via a current preamplifier in the external circuit for demodulation. By monitoring photocurrent values at 0 mV bias, the excitation of the localised surface plasmon resonance (LSPR) in the nanogap of the tunnelling junction can be fine-tuned.

Single-protein conductance measurements via switch strategy.

For single-protein 'switch' measurements, WDR5 and Hsp90 were measured. The WDR5 protein with an N-terminal 6×His tag and a mono-streptavidin (MSA) tag at the C-terminus enables two distinct strategies for constructing single-protein switch configurations. In the first strategy, the MSA tag allows for site-specific binding to the biotin-modified tunnelling electrode surface. Biotin PEG thiol (0.1 mM) was reduced in 10 mM TCEP for 1 hour to expose the -SH bond, then linked to the tunnelling electrode for more than 4 hours. The functionalized tunnelling electrode was then rinsed with DI water to remove residual biotin. Subsequently, WDR5 (30 μg/mL) was introduced to the flow cell for 30 minutes and allowed the MSA tag to bind to the biotin-modified tunnelling electrode, anchoring the protein to the tunnelling junction via one end. After incubation, the flow cell was flushed with 1 mL PBS (1 mM) buffer to exclude any residual proteins, and the flow cell was mounted on the experimental setup for conductance measurements in PBS. Alternatively, the N-terminal His-tag of WDR5 can be immobilised on the tunnelling electrode via Cu-NTA coordination. C2-NTA (1.3 mM) was reduced in 100 mM TCEP for 1 hour and then linked to the tunnelling electrode for more than 4 hours. After modifying NTA, the tunnelling probe in the flow cell was rinsed four times with DI water. Then the NTA-functionalized tunnelling probe was incubated with 5 mM $CuSO_4$ for 10 minutes to chelate the NTA groups. Excess copper ions were removed by washing with PBS. WDR5 was then introduced to the flow cell for 30 minutes and allowed the His-tag to coordinate with $Cu^{2+}$-NTA, followed by PBS washing. Finally, the flow cell was subsequently mounted for conductance measurements in PBS. Hsp90, which carries an N-terminal 6×His tag, was functionalized using a similar protocol as described for the His-tagged WDR5.

Electrical tracking of Hsp90 conformational dynamics

For conformational kinetic analysis, the Hsp90-functionalized tunnelling probe was prepared and measured through the following steps: (1) The Hsp90-functionalized QMT probe was rinsed with PBS (1 mM, pH 7.2–7.4), and 40 μL of PBS was added to the flow cell as the testing buffer. In the absence of illumination, the His-tag tethering Hsp90 to the electrode surface behaves as a flexible, long-chain linker, allowing the Hsp90 molecule to remain suspended in solution with only occasional contact with the opposing electrode, resulting in a low probability of current jump signals. (2) A laser beam was then focused onto the tip of the tunnelling electrode to excite the LSPR effect, generating an optical gradient force directed toward the nanogap. Under optical force, suspended Hsp90 was drawn toward the nanogap, thereby bridging the tunnelling junction and resulting in an overall increase in tunnelling current. (3) While the laser remains on and Hsp90 bridges the nanogap, ATP (1 mM) is introduced into the flow cell, and the current trace is recorded in real-time for 20 minutes to monitor the interaction between Hsp90 and ATP. (4) After recording the ATP-induced conformational dynamics, the remaining ATP in the solution is flushed out with fresh PBS buffer, and AMP-PNP (0.5 mM), a non-hydrolyzable ATP analogue known to lock Hsp90 in a closed conformation, is introduced into the flow cell. (5) After finishing both ATP and AMP-PNP experiments, 1 mM EDTA is added to the flow cell to disrupt the coordination between the His-tag and the $Cu^{2+}$ ions at the tunnelling electrode. As a result, EDTA, a common chelating agent for metal ions, effectively releases His-tagged proteins from the electrode surface. After incubating the Hsp90-functionalized QMT probe in EDTA for several minutes, the electrode is rinsed with buffer to remove residual EDTA and dissociated His-tagged protein. Finally, the QMT probe is measured again under both dark and illumination conditions, following the same protocol as in the plasmonic optical trapping experiments, to validate the successful dissociation of Hsp90 from the electrode.

Data analysis

Data analysis was performed using custom MATLAB scripts. OriginLab was used for statistical evaluation. I–V curves were analysed using the Simmons model to extract tunnelling parameters: gap distance (d), active tunnelling area (A), and barrier height ($\phi_B$). Dwell time distributions were fitted with single exponential decay functions. Capture probability was estimated using a non-parametric bootstrapping method.

Finite-difference time-domain simulations

The electric field intensity distribution and absorption/scattering spectra were simulated using the finite-difference time-domain method (Lumerical: ANSYS, Inc.). A model was constructed containing two spherical gold nanoparticles, each with a diameter of 50 nm. Additionally, we incorporated protruding structures with a curvature radius of 5 nm between the nanoparticles to simulate the irregular surface structure of the nanoelectrodes formed by feedback electroplating. The gap between the gold nanoparticle pair was set to 0.5 nm. For boundary conditions, a perfect matching layer (PML) was employed, ensuring that the entire nanostructure was fully enclosed within the simulation region. The refractive index of the medium in the simulation region was set to 1.33 to mimic a water environment. In the simulation, we selected a total-field scattered-field (TFSF) source, with the source intensity set to 1 mW/$\mu m^2$, incident along the Z-axis. The light source structure fully encapsulated the simulated nanostructure. An automatic non-uniform grid division was applied across the entire simulation region. Furthermore, to ensure simulation accuracy, the grid size was set to 5 nm in the region encompassing the entire nanostructure model, and 0.1 nm in the region containing the nanoparticle gap and protruding structures (20 × 20 nm in length and width). The refractive index distribution clearly shows the sharp edges of the nanostructures, indicating that the grid division is reasonable and effective. We performed a frequency-domain calculation to simulate the scattering, absorption, and extinction cross-sections of the constructed structure. The optical force was derived using the dipole approximation model for the liquid crystal molecule 5CB. The molecular polarizability of 5CB was calculated using density functional theory (DFT) with Gaussian16 (B3LYP and 6-311 basis set), considering the effects of the directional electric field induced by the applied DC bias during tunnelling measurements.

**Supplementary Text**

Single-molecule events analysis

***Baseline fitting through asymmetric least squares smoothing.*** The baseline is estimated via the asymmetric least squares (ALS) method[2] by integrating asymmetric weighted residuals with smoothing constraints. Its core principle involves assigning reduced weights to signal peaks, which lie above the baseline, and increased weights to data points near the baseline. This approach ensures the baseline conforms to the signal's lower regions during iterative fitting. The detailed procedure is as follows:

The original signal is $\boldsymbol{y} \in \mathbb{R}^n$. A downsampled sub-signal $\boldsymbol{yy} \in \mathbb{R}^m$ is generated by selecting every step-th sample from $\boldsymbol{y}$, where $\lfloor m = n/step \rfloor$. The baseline estimation is performed on $\boldsymbol{yy}$, and the final baseline is interpolated to match the original signal length $n$.

For the downsampled signal $\boldsymbol{yy}$, the objective function combines asymmetric weighted residuals and smoothness constraints:

$$J(z) = \sum_{i=1}^{m} \omega_i (yy_i - z_i)^2 + \lambda \sum_{i=2}^{m-1} (z_{i-1} - 2z_i + z_{i+1})^2 \quad (1)$$

Where, $\omega_i$ is the asymmetric weight, $\lambda$ is the smoothing parameter.

$$\omega_i = \begin{cases} p, & \text{if } yy_i > z_i \\ 1-p, & \text{if } yy_i < z_i \end{cases} \quad (2)$$

The matrix form of the objective function is:

$$J(\mathbf{z}) = (\mathbf{yy} - \mathbf{z})^T \mathbf{W} (\mathbf{yy} - \mathbf{z}) + \lambda \mathbf{z}^T \mathbf{D}^T \mathbf{D} \mathbf{z} \quad (3)$$

Where $\boldsymbol{W}$ is a diagonal matrix with elements $\omega_i$. Taking the derivative with respect to $\boldsymbol{z}$ and setting it to zero yields the linear system:

$$(\mathbf{W} + \lambda \mathbf{D}^T \mathbf{D}) \mathbf{z} = \mathbf{W} \mathbf{yy} \quad (4)$$

Initialise the baseline $z^{(0)} = yy$, set initial weights $\omega_i{}^{(0)} = 1$. Fixed ten iterations:

Step 1: Construct the diagonal weight matrix $\boldsymbol{W}^{(k)}$:

$$\omega_i{}^{(k)} = p \cdot \mathbb{I}(yy_i > z_i) + (1-p) \cdot \mathbb{I}(yy_i < z_i) \quad (5)$$

Where $\mathbb{I}(\cdot)$ is the indicator function.

Step 2: Solve the linear system using Cholesky decomposition:

$$\mathbf{C} = \mathrm{chol}\left(\boldsymbol{W}^{(k)} + \lambda \boldsymbol{D}^T \boldsymbol{D}\right), \ \mathbf{z}^{(k+1)} = \mathbf{C}^{-1}(\mathbf{C}^{-T}(\mathbf{W}^{(k)} \mathbf{yy})) \quad (6)$$

Step 3: Update the weights $\omega_i{}^{(k+1)}$.

The estimated baseline $\boldsymbol{z} \in \mathbb{R}^m$ is interpolated to the original length $n$ using linear interpolation, yielding the final baseline $\boldsymbol{zz} \in \mathbb{R}^n$.

In ALS baseline fitting, the key parameters are step, p, and λ, and the smoothness of the baseline is controlled by adjusting these three parameters.

***Recognition of signal peaks.*** The background current is assumed to comply with a Gaussian distribution, with its probability density function defined as follows:

$$f(x) = \frac{1}{\sigma\sqrt{2\pi}} \cdot e^{-\frac{(x-\mu)^2}{2\sigma^2}} \quad (7)$$

Where $\mu$ and $\sigma$ are the mean and standard deviation of the Gaussian distribution, respectively.

Then, a dual threshold strategy is set for signal recognition. The primary threshold is to determine the signal edge, which is defined as:

$$\mathbf{Th_1} = \mathbf{zz} + S_1 * \sigma \quad (8)$$

Where $\boldsymbol{zz}$ is the baseline, $\sigma$ is the standard deviation of Gaussian fitting, $S_1 \in \mathbb{R}$. When the current value of the data in $\boldsymbol{y}$ is greater than $\boldsymbol{Th_1}$, it is considered a possible signal point.

The second level threshold includes $\boldsymbol{Th_2}$ and $\boldsymbol{T_{th}}$:

$$\mathbf{Th_2} = \mathbf{zz} + S_2 * \sigma \qquad (9)$$
$$T_{th} = 0.01 \text{ ms} \qquad (10)$$

Where $\boldsymbol{zz}$ is the baseline, $\sigma$ is the standard deviation of Gaussian fitting, $S_2 \in \mathbb{R}$. $\boldsymbol{Th_2}$ and $\boldsymbol{T_{th}}$ are defined to further screen the signal. When the average height of the signal is greater than $\boldsymbol{Th_2}$ and its width is greater than $T_{th}$, it is considered a valid signal. Based on the threshold above, data points in the original signal $\boldsymbol{y}$ can be labelled as "signal peak" and "noise". Meanwhile, the starting and ending points of each signal peak are recorded.

***Statistical analysis of signal peaks.*** Each signal peak identified in the previous step is designed as an "event". For each event, expand its extended front and back edges (the starting and ending points) by a specified number of points, $L$ (usually, $L = 50$), to obtain more complete event information. Ensure that all data points in the extended area are categorised as the "baseline". Otherwise, discard them and define the extended area as the local baseline for the corresponding event. All events are temporally aligned based on their timeline midpoints. Construct a two-dimensional probability distribution of events using the kernel density estimation method and perform histogram analysis on the amplitude dimension. The optimal histogram binning is determined through the Scott criterion. Additionally, two characteristic parameters, amplitude and dwell time, are calculated for each event. Amplitude is the average current difference between the start and end points of the signal peak and the local baseline, while dwell time is the duration of the signal peak from the start to the end point.

***K-means Clustering analysis.*** The goal of k-means clustering is to assign $n$ data points to $k$ clusters[3], which can be characterised by the following objective function[4]:

$$J = \sum_{j=1}^{k} \sum_{x \in C_j} \| x - \mu_j \|^2 \qquad (11)$$

Where $C_j$ represents the cluster $j$, $\mu_j$ is the centroid of cluster $j$, $\| x - \mu_j \|^2$ is the Squared Euclidean distance.

Firstly, randomly select $k$ initial centroids as initial clusters. Calculate the distance between each data point and $k$ centroids in sequence, assign it to the centroid with the closest distance, and then update the centroid of this cluster. Continuously iterate the above process until all $k$ centroids no longer change.

The clustering results can be evaluated by internal and external indicators. External indicators are used to evaluate the accuracy of data classification when the true labels are known. Internal indicators are used to characterise the quality of clustering results by calculating the intra-class cohesion and inter-class dispersion. The following indicators are commonly used[5].

Calinski-Harabasz value:

$$CH_k = \frac{S_B}{S_W} \times \frac{(N-k)}{(k-1)} \qquad (12)$$

Where $S_B$ and $S_W$ are the total inter-cluster and intra-cluster variances, $k$ is the number of clusters, $N$ is the number of observations. The greater the inter-cluster variance and the lower the intra-cluster variance, the larger the $CH_k$, which means better separation between clusters.

Davies-Bouldin value:

$$DB_k = \frac{1}{k} \sum_{i=1}^{k} \max_{j=1\sim k, j \neq i} \left( \frac{W_i + W_j}{C_{ij}} \right) \qquad (13)$$

$k$ is the number of clusters, $W_i$ is the distance from all sample points in $C_i$ to the centroid $C_i$, $W_j$ is the distance from all sample points in $C_j$ to the centroid $C_j$, and $C_{ij}$ is the distance from the centroid $C_i$ to $C_j$. The smaller $DB_k$ means the lower similarity between clusters, thus the better clustering result.

Silhouette Coefficient:

$$\mathrm{Sil(i)} = \frac{\mathrm{b_i - a_i}}{\mathrm{max(a_i, b_i)}} \quad (14)$$

$a_i$ is the distance from $i$ to all other sample points in the same cluster, and $b_i$ is the distance from $i$ to all other sample points in the nearest neighbouring cluster. The value range of $Sil(i)$ is [-1,1]. The clustering is excellent if $Sil(i)$ approaches 1, which indicates that the intra-class difference is much less than the smallest inter-class difference.

Optimising the plasmonic trapping efficiency with photocurrent

The photocurrent arises from a combination of different mechanisms, including optical rectification, thermal expansion, photothermoelectric effects, and plasmon-induced hot electron generation.[6-12] Among these, contributions from optical rectification and plasmonic hot electron effects are dictated by LSPR and show polarisation dependence on the incident laser. Therefore, by rotating polarisation angles, the incident light's polarisation can be aligned with the LSPR coupled direction, maximising local electromagnetic field enhancement and improving plasmonic trapping efficiency. Once the LSPR excitation was optimised, the light source was switched to a DC-driven mode, and the signal acquisition was transferred to the patch-clamp system for tunnelling current detection. Tunnelling sensing of target molecules was then performed without/with light activation. Under laser illumination, the laser intensity is initially set to 500 μW/μm$^2$ and gradually increased while recording the current-time traces in real-time. When a target molecule diffused into the tunnelling sensing region between the electrodes in solution, a current jump was observed. The frequency and duration of these current jumps reflect the occurrence and residence time of individual molecules in the sensing region. In dark conditions, molecules undergo Brownian motion and enter the sensing region sporadically with low frequency and short dwell times. Conversely, under laser illumination, the excitation of LSPR on the gold nanoelectrodes generates an optical gradient force directed toward the sensing region, actively trapping molecules in the vicinity of the nanogap. This results in more frequent and prolonged current jumps in current-time traces. When the optical gradient force is sufficiently strong, it can stably confine a molecule within the tunnelling region, leading to a sustained current level in the time trace.

The estimation of optical force

The optical force was derived using the dipole approximation model for single molecules. In the localised hotspots of the nanogap, the force acting on the trapped molecules is the sum of the optical gradient force, $\mathbf{F}_g = 0.5\alpha'\nabla|\mathbf{E}|$, and the optical scattering force, $\mathbf{F}_s = 0.5k\alpha''\nabla|\mathbf{E}|$, where $\mathbf{E}$ is the electric field and $\boldsymbol{k}$ is the wave vector. Here, $\alpha$ is the polarizability of the particle, while $\alpha'$ and $\alpha''$ represent the real and imaginary parts of the polarizability, respectively. In plasmonic nanostructures, the gradient force is directed toward the centre of the gap, i.e., the position of the strongest local hotspot of the electromagnetic field. The scattering force arises from the transfer of photon momentum and acts in the direction of light propagation. In the overall model, the laser is incident along the Z-axis, directly facing the tip of the tunnelling probe, and the scattering force is also directed toward the nanogap. However, for non-resonant molecules ($\alpha' >> \alpha''$), the optical gradient force is much larger than the optical scattering force ($\mathbf{F}_g \geq \mathbf{F}_s$), and the scattering force's effect on small molecule trapping can be neglected.

For anisotropic molecules with sizes in the sub-10 nm range, the form of molecular polarizability is more complex and can be calculated using density functional theory (DFT) with Gaussian16 (B3LYP and 6-311 basis set). For optical force calculations, the liquid crystal molecule 5CB was used as a model, while also considering the effects of the directional electric field induced by the applied DC bias during tunnelling measurements. The three-dimensional polarizability tensor of 5CB was denoted as

$$\begin{array}{lll}(7.77925e-39, & -3.77294e-40, & -5.32162e-40, \\ -3.77294e-40, & 4.32357e-39, & -3.83858e-40, \\ -5.32162e-40, & -3.83858e-40, & 3.44880e-39,)\end{array}$$

, representing the molecular polarisation response in different directions. For the optical force calculation within the XY-plane, the elements of the polarizability tensor, $\alpha_{xx}$, $\alpha_{xy}$, $\alpha_{yx}$, and $\alpha_{yy}$, which represent the polarisation responses of the molecule along the X and Y directions, were used. The elements of the polarizability tensor, $\alpha_{xx}$, $\alpha_{xz}$, $\alpha_{zx}$, and $\alpha_{zz}$, were used to calculate the optical force within the XZ-plane. Furthermore, we computed the trapping potential ($\mathbf{U}=-0.5\alpha|\mathbf{E}|^2$) and normalised it with the thermal energy $kT_0$, where $k$ is Boltzmann's constant and the temperature $T_0 = 300\ K$.

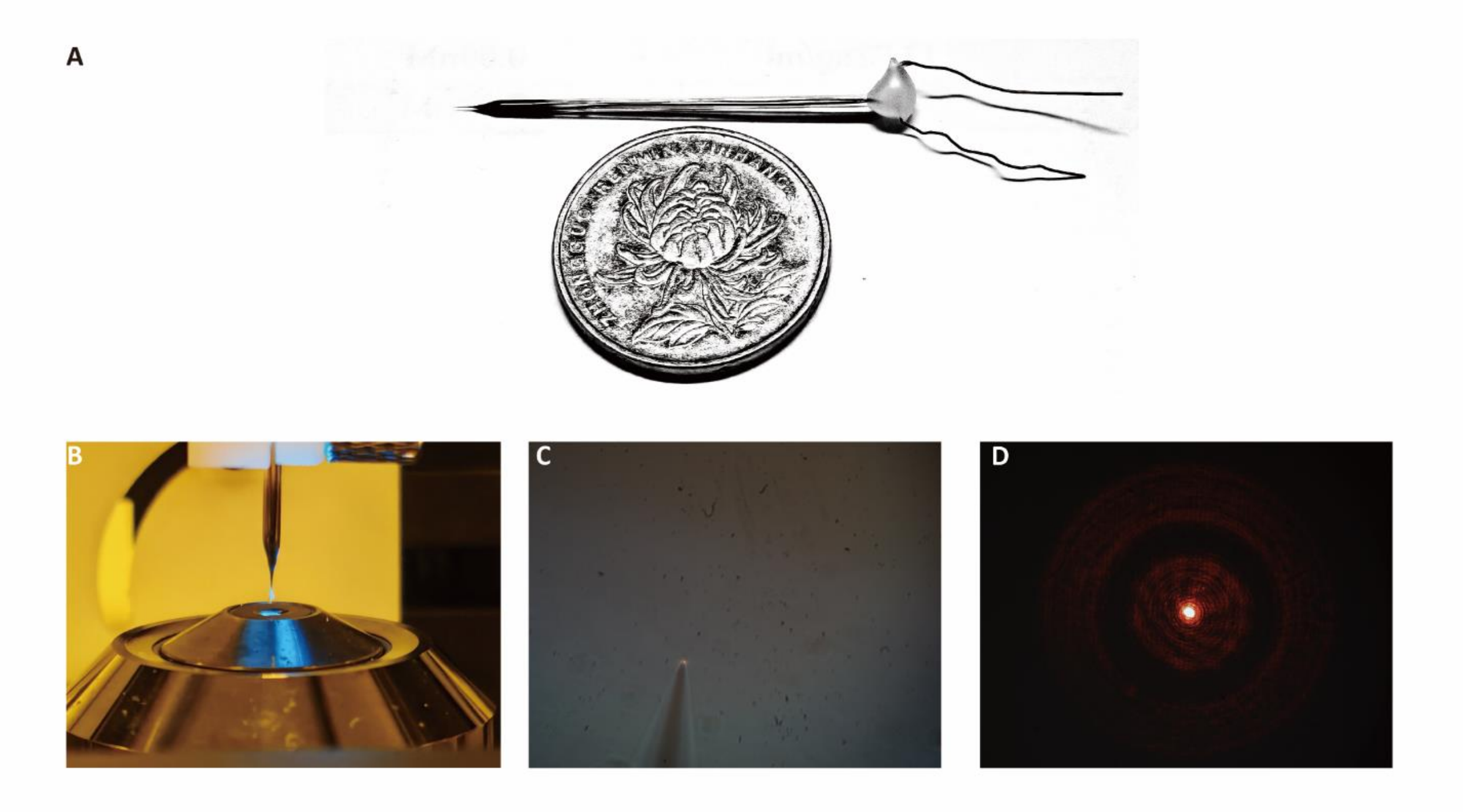

**Fig. S1.**

Type or paste caption here. Create a page break and paste in the figure above the caption. Optical images of the QMT probes. Optical images of **(A)** the whole QMT probe and **(B)** the QMT probe exposed to laser irradiation in air. The bright-field images of **(C)** the QMT probe under laser irradiation and LED illumination in solution, and **(D)** the QMT probe in solution when the light is focused on the tip, showing the diffraction rings.

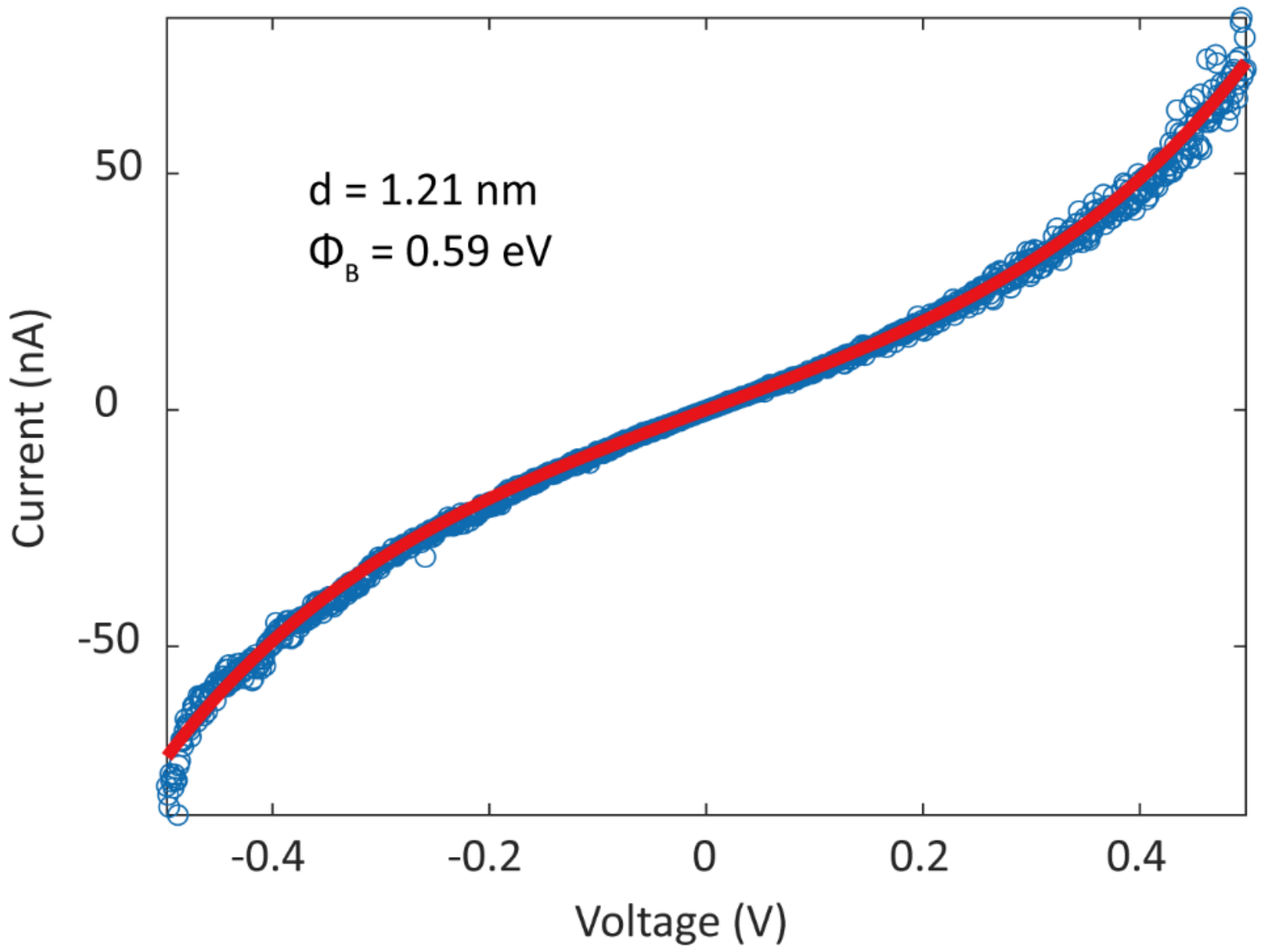

**Fig. S2.**
**The current-voltage trace was recorded in DI water and fitted with the Simmons model.**

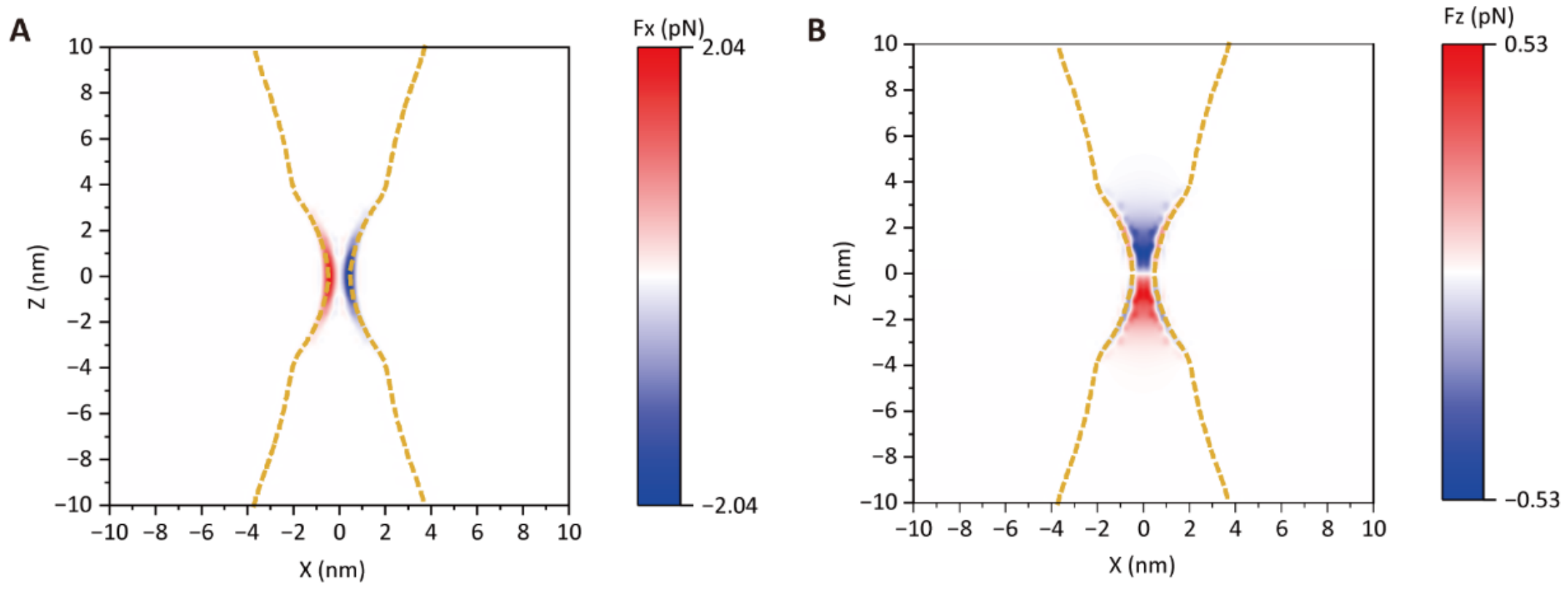

**Fig. S3.**

**The calculation of the gradient optical force exerted on 5CB at the X-Z plane. (A)** The x component of the gradient optical force and **(B)** z component of the gradient optical force show that the forces are directed towards the electrode surface in the gap region. The incident light is along the Z-axis, and the polarisation is parallel to the X-axis. Wavelength: 637 nm, gap distance: 0.5 nm, Incident intensity: 1 mW/μm².

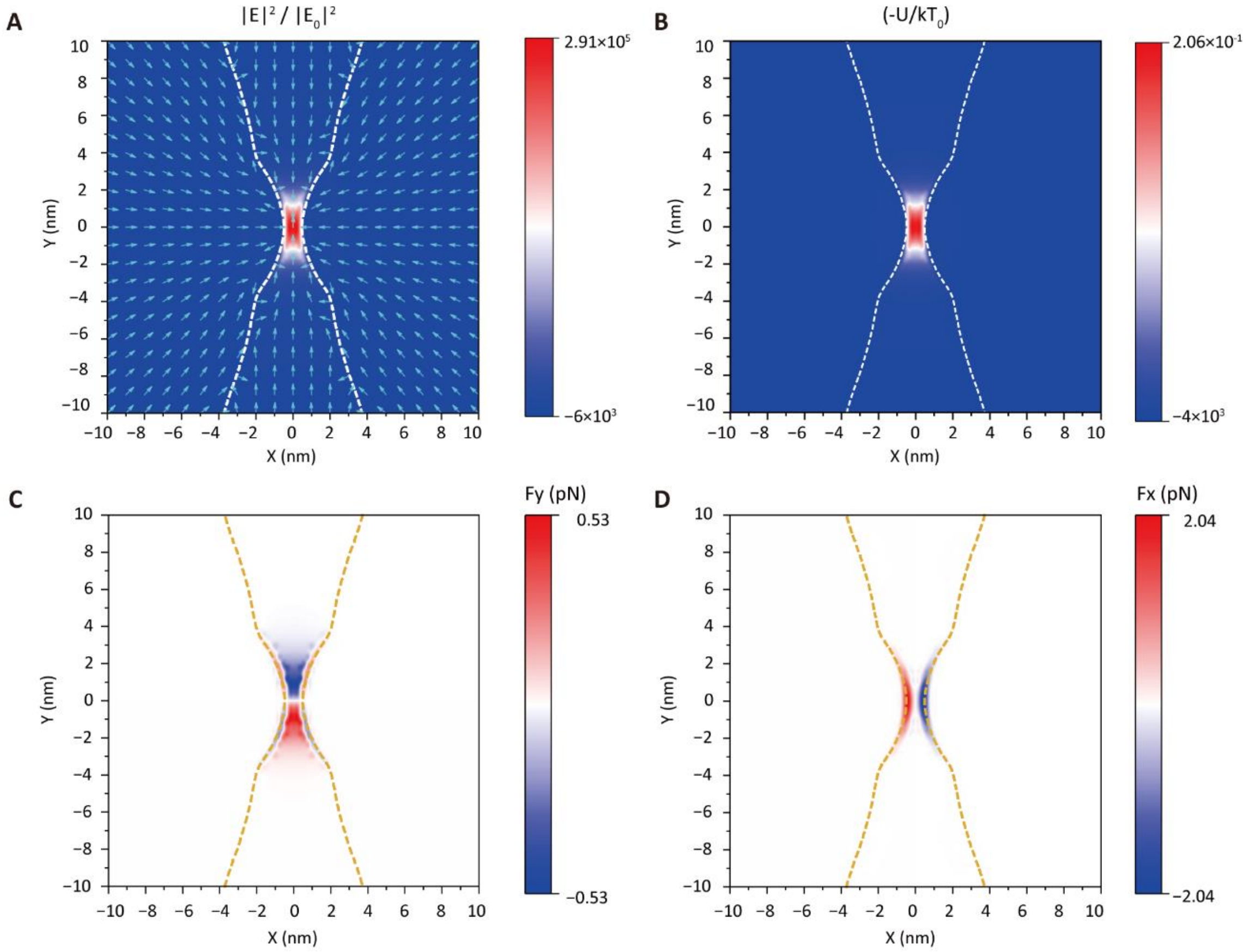

**Fig. S4.**

**The calculated spatial distribution of the field enhancement factor, trapping potential cross-section, and gradient optical force exerted on 5CB at the X-Y plane. (A)** The spatial distribution of the field enhancement, together with the sparse optical force vector, shows a plasmonic hotspot located at the tunnelling junction. **(B)** The trapping potential cross-section in units of $kT_0$ ($T_0$ = 300 K). **(C)** The y component and **(D)** x component of the gradient optical force. The incident light is along the Z-axis, and the polarisation is parallel to the X-axis. Wavelength: 637 nm, gap distance: 0.5 nm, Incident intensity: 1 mW/µm².

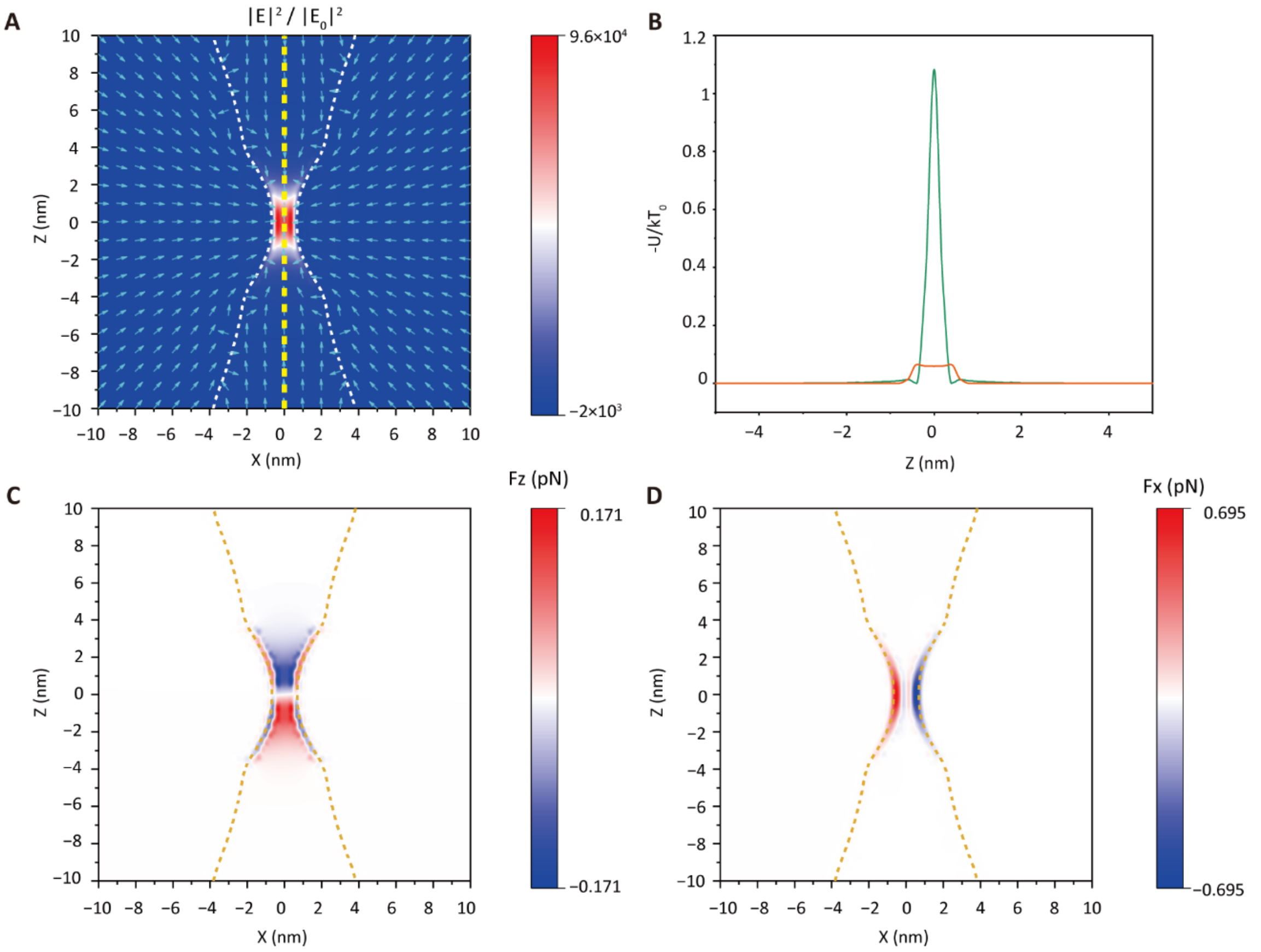

**Fig. S5.**

**Comparison of the calculation results with different gap distances. (A)** The calculated spatial distribution of the field enhancement factor, together with the sparse optical force vector, with a gap distance of 1 nm. **(B)** Comparison of trapping potential with the gap distances of 0.5 nm and 1 nm. **(C)** The z component and **(D)** x component of the gradient optical force with a gap distance of 1 nm. The incident light is along the Z-axis, and the polarisation is parallel to the X-axis. Wavelength: 637 nm, Incident intensity: 1 mW/μm², X-Z plane, 5CB.

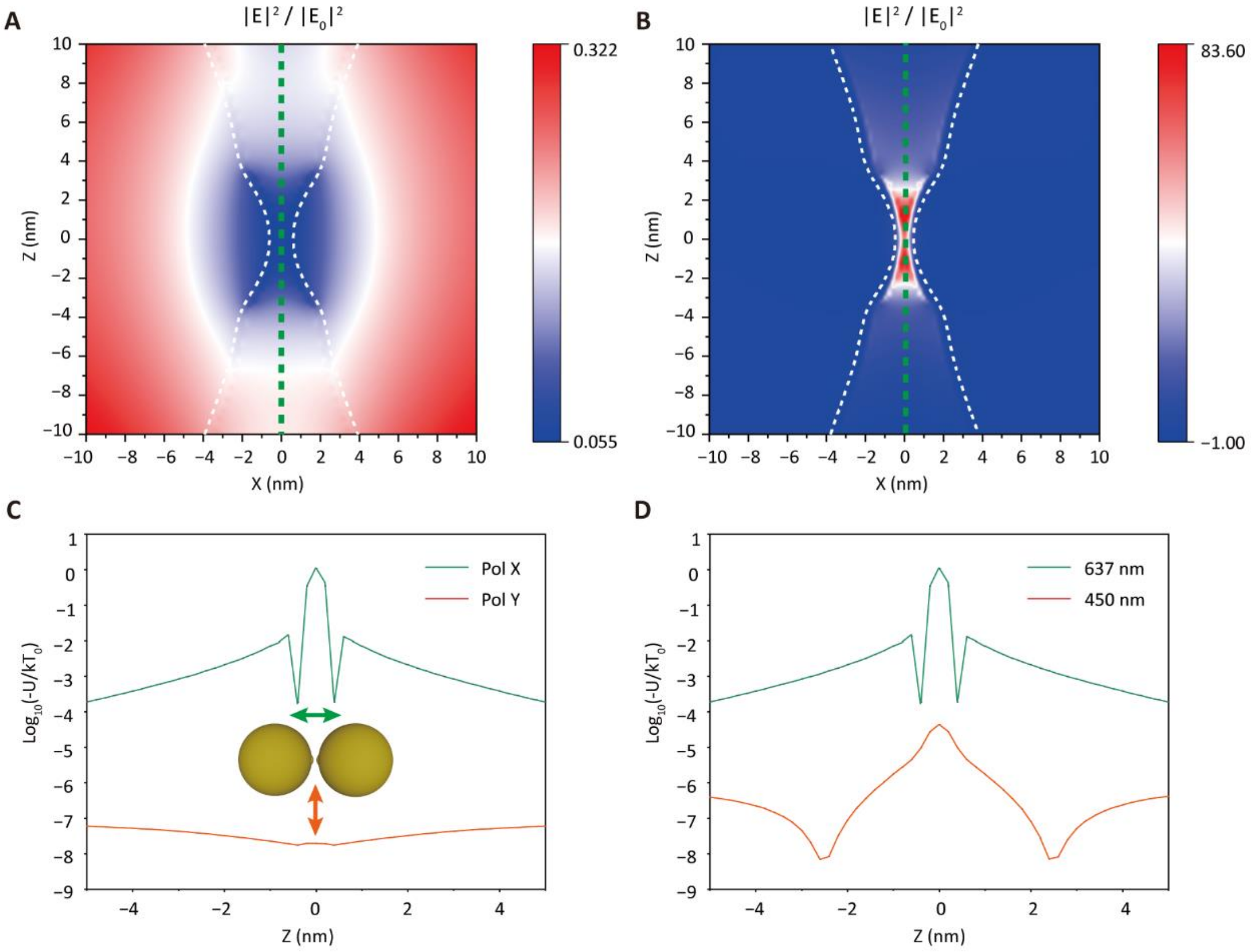

**Fig. S6.**

**The calculation results of polarisation and wavelength dependence. (A)** The calculated spatial distribution of the field enhancement with polarisation parallel to the Y direction. **(B)** The calculated spatial distribution of the field enhancement under 450 nm illumination. **(C)** Comparison of tapping potential in logarithmic scale with the polarisation parallel (the X direction) and perpendicular (the Y direction) to the coupling direction. **(D)** Comparison of tapping potential in logarithmic scale under 637 nm and 450 nm illumination. The incident light is along the Z-axis. Incident intensity: 1 mW/μm², X-Z plane, 5CB.

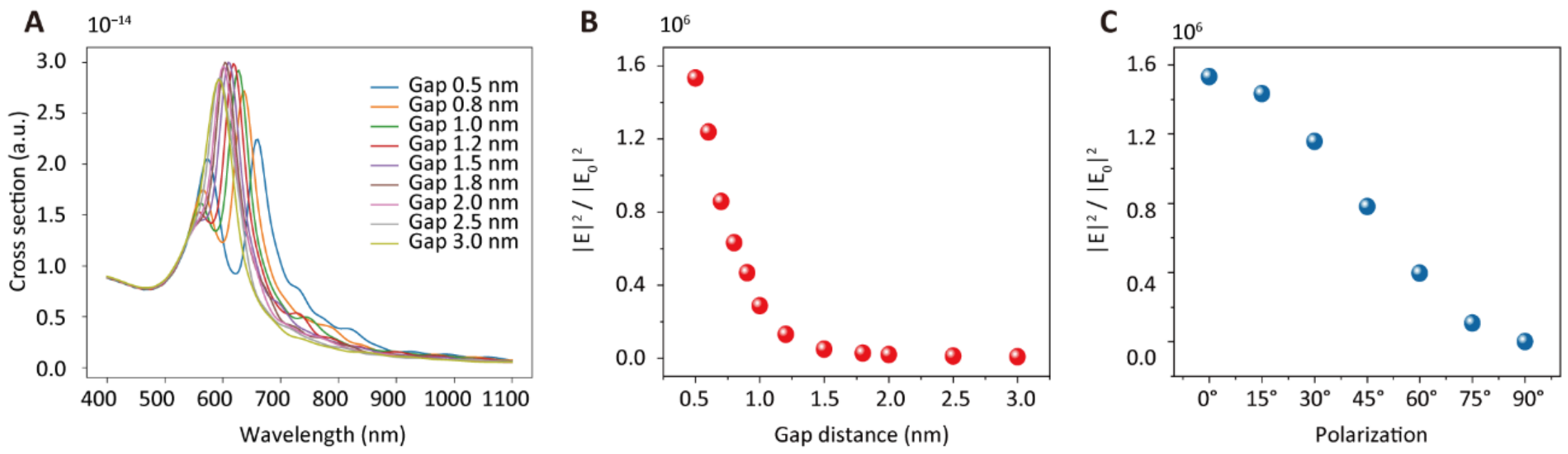

**Fig. S7.**

**The calculated field enhancement factor is dependent on the gap distance and polarisation.** **(A)** The calculated extinction spectra of a tunnelling junction with varied gap distances. The calculation results show that the theoretical resonance wavelength of the plasmonic tunnelling junction is approximately the operated 637 nm, matching the measured scattering spectra. Besides, the resonance peak shifts blue as the gap distance narrows. **(B)** The plot of the field enhancement factor versus gap distance. **(C)** The plot of the field enhancement factor versus polarisation. The incident light is along the Z-axis. Wavelength: 637 nm, Incident intensity: 1 mW/μm².

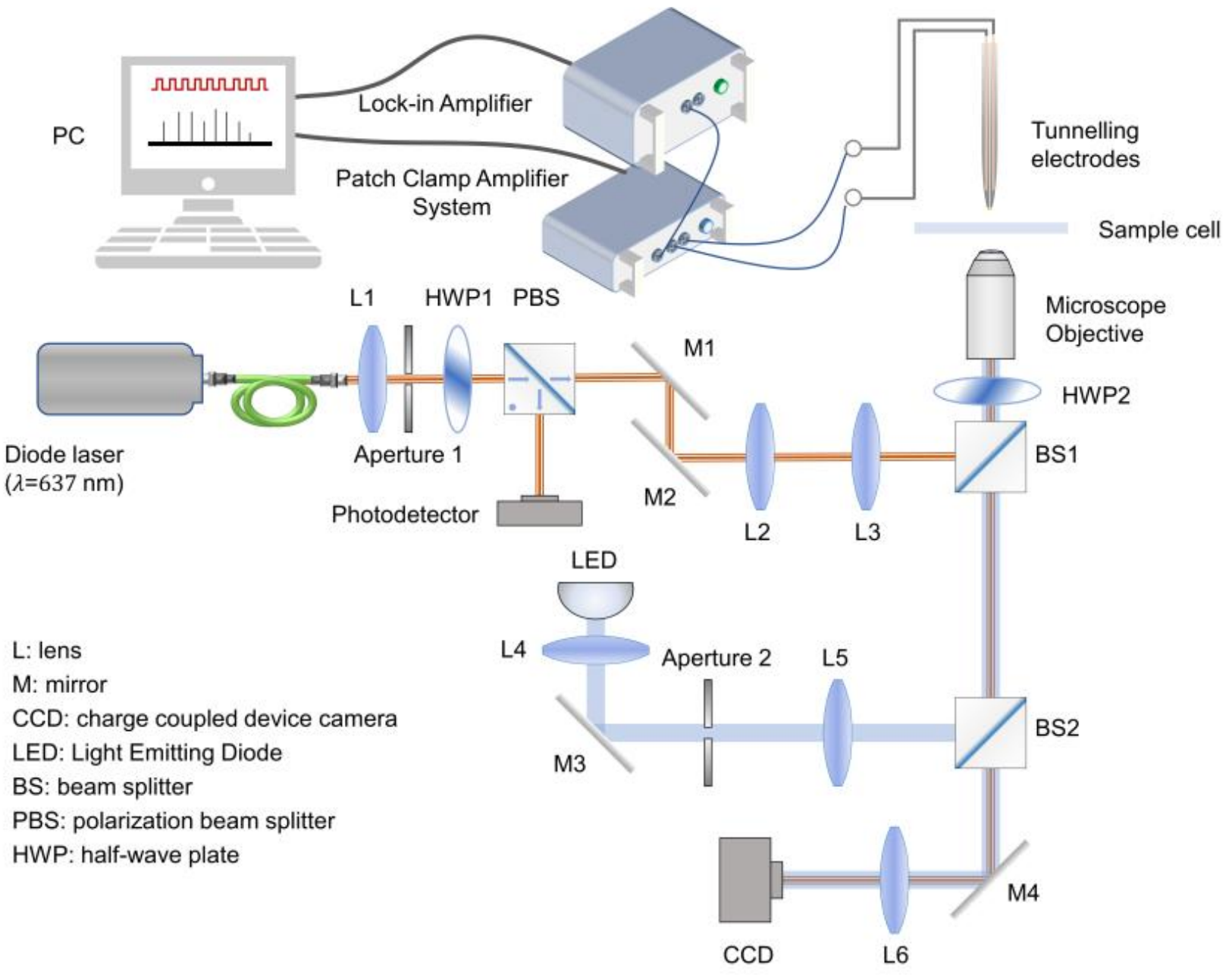

**Fig. S8.**

**Experimental apparatus of single-molecule plasmonic tunnelling nanoscopy.**

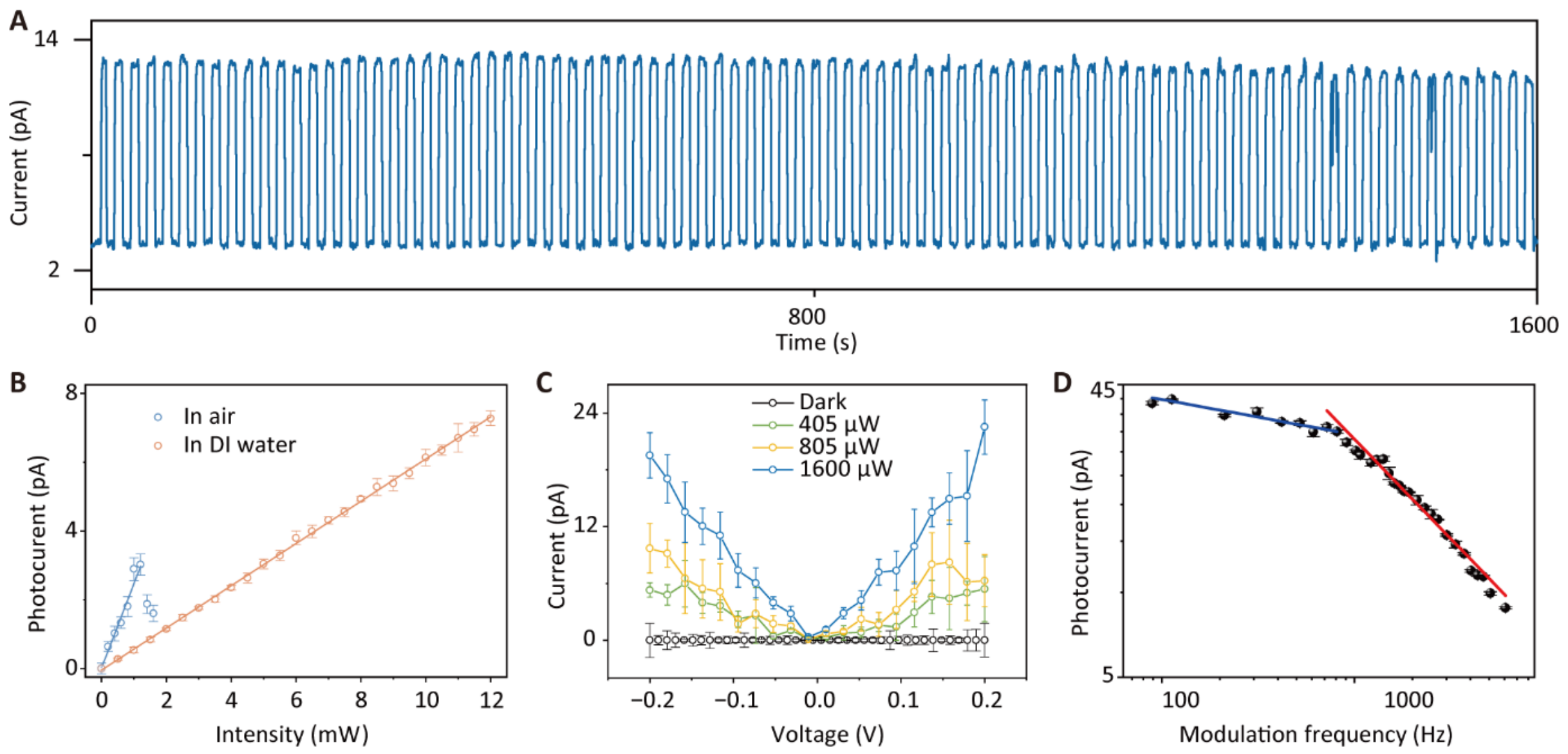

**Fig. S9.**

**Characterisation of photocurrent generation at QMT probes. (A)** Temporal response of the photocurrent under alternating illumination conditions. The laser was periodically modulated (on/off cycles) over a duration of 25 minutes, demonstrating stable and reversible photocurrent generation. **(B)** Plot of photocurrent versus laser power. The photocurrent measured in air (blue circle) was higher than that measured in deionised water (orange circle), while the QMT probes exhibited greater stability in deionised water. **(C)** Plot of photocurrent versus applied voltage. **(D)** Plot of photocurrent versus modulation frequency of excitation light.

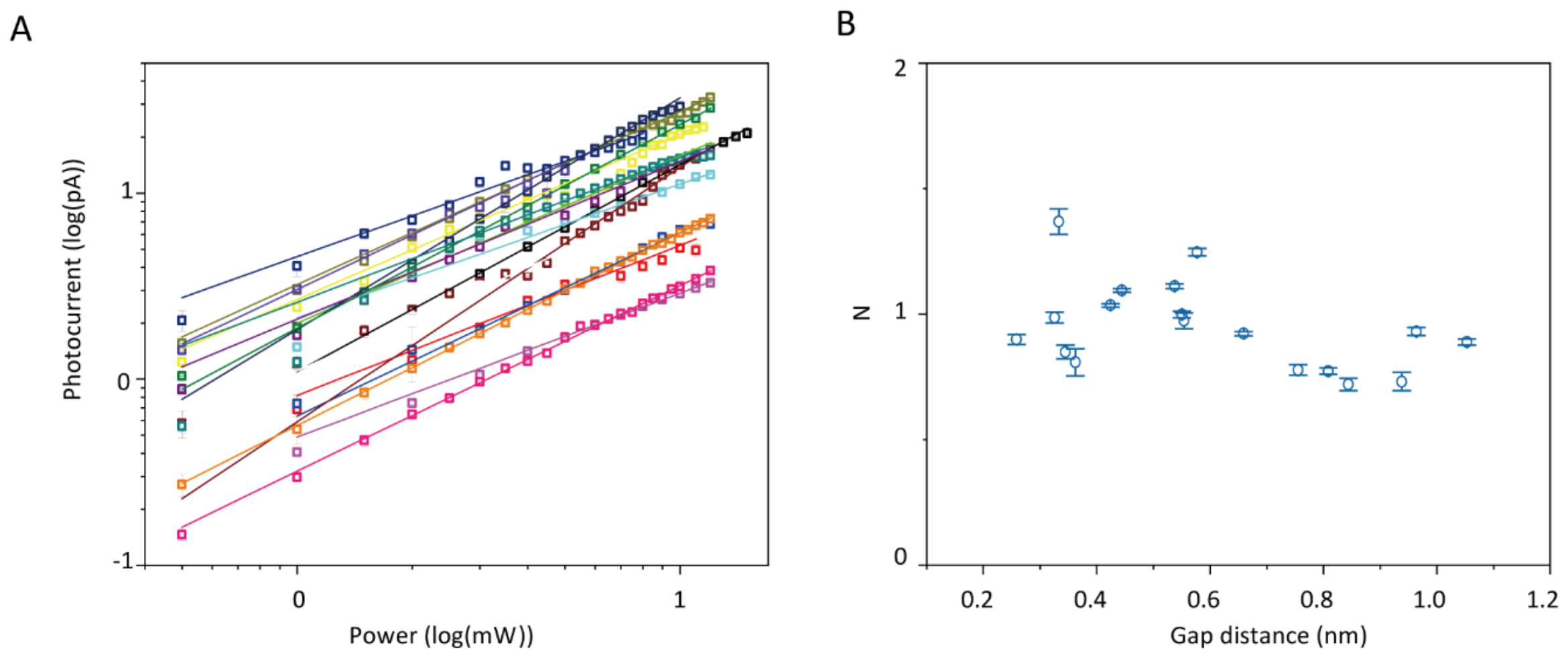

**Fig. S10.**

**The power dependency of the photocurrent for different devices. (A)** The plot of photocurrent versus light power in a double logarithmic scale with linear fitting (solid lines). **(B)** The plot of the slope (the power-law exponent, N) of the linear fitting versus the gap distance for the measured QMT probes indicates that a one-order power-law scaling corresponds to one-photon absorption.

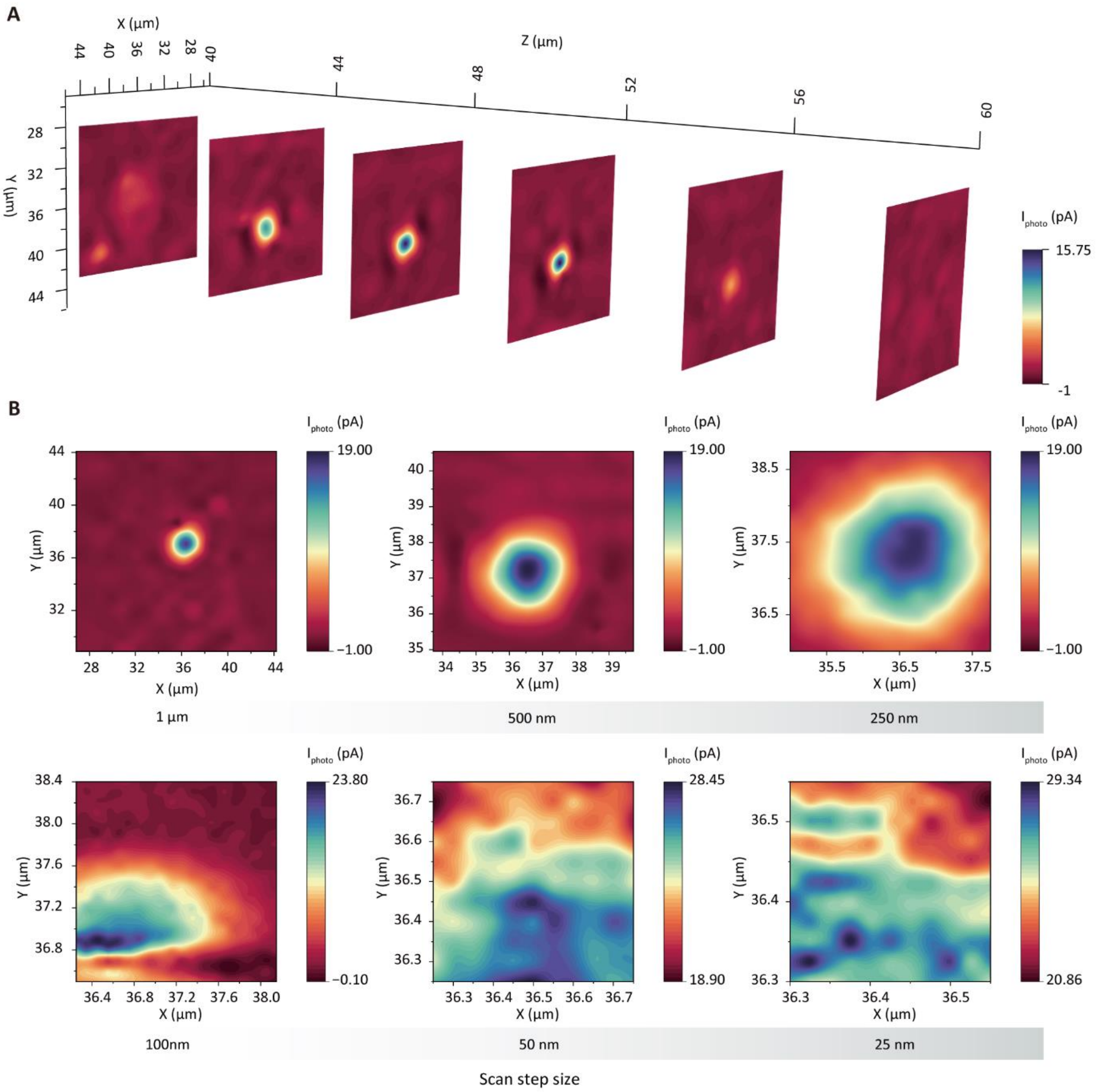

**Fig. S11.**

**Spatial mapping of photocurrent response at QMT probe apex. (A)** Three-dimensional photocurrent mapping around the apex of the QMT probe. The scan step sizes along the X, Y, and Z axes were 1 μm, 1 μm, and 4 μm, respectively. **(B)** Two-dimensional photocurrent scanning in the XY-plane. Step sizes were progressively reduced from 1 μm to 25 nm, showcasing the device's nanoscale spatial resolution capabilities. Modulation frequency: 1077 Hz, laser power: 1.1 mW, bias: 0 mV, in air.

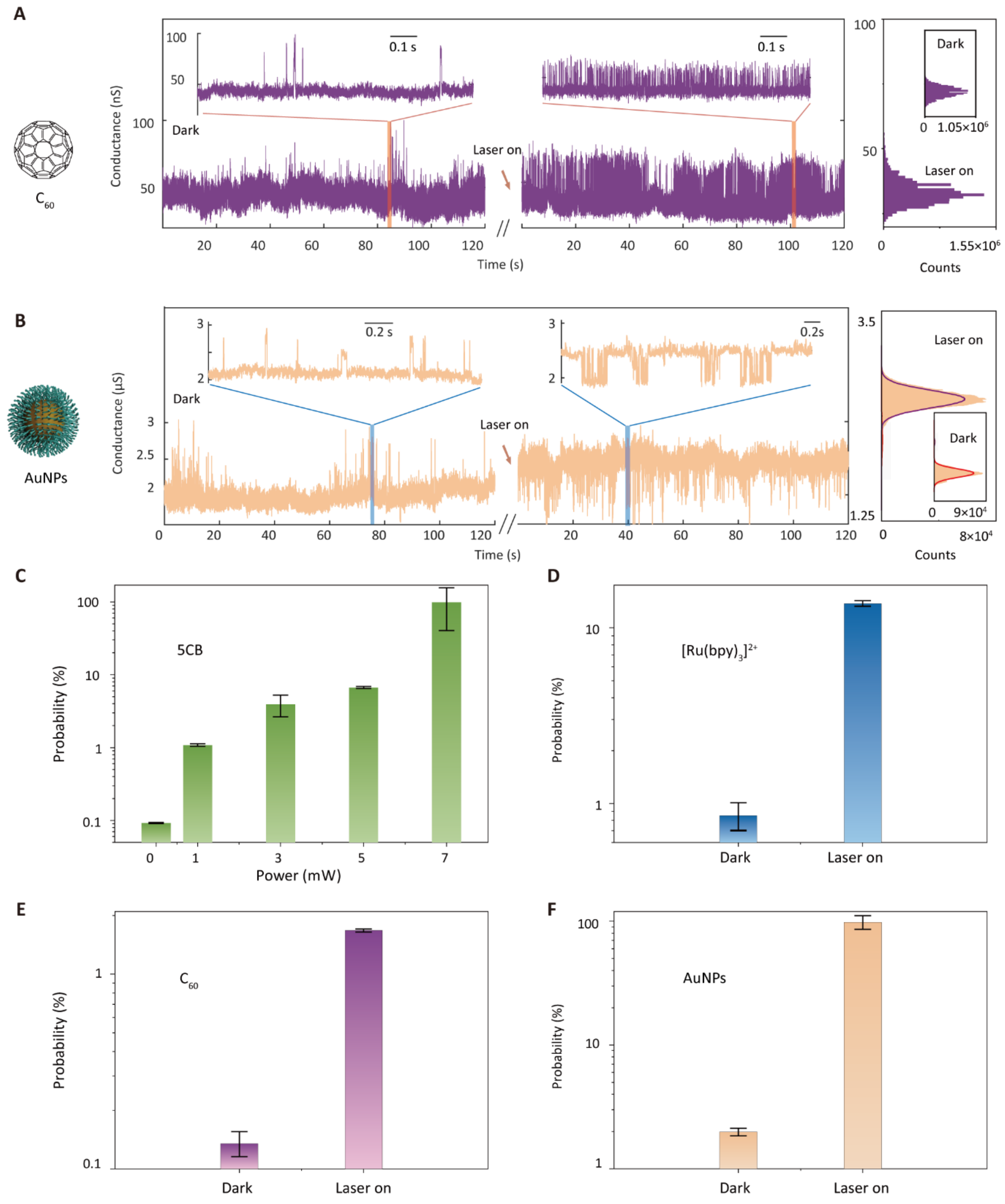

**Fig. S12.**

**Plasmonic trapping and conductance detection of single molecules of varying sizes. (A-B)** Representative conductance traces and histograms of **(A)** $C_{60}$ fullerenes at 100 mV and **(B)** AuNPs recorded at a 100 mV bias under laser-off and laser-on conditions. **(C-F)** Quantitative analysis of the capture probability for **(C)** 5CB at different laser intensities, **(D)** $[Ru(bpy)_3]^{2+}$, **(E)** $C_{60}$, and **(F)** AuNPs. Capture probabilities were calculated by extracting peak events from conductance traces and fitting the dwell time distributions. Data are plotted on a logarithmic scale. Each data point denotes the mean capture probability, calculated by bootstrapping ($N_{boots}$ = 2000). Error bars represent the 90% confidence interval from the bootstrapped mean value of capture probability (fig. S30, S31).

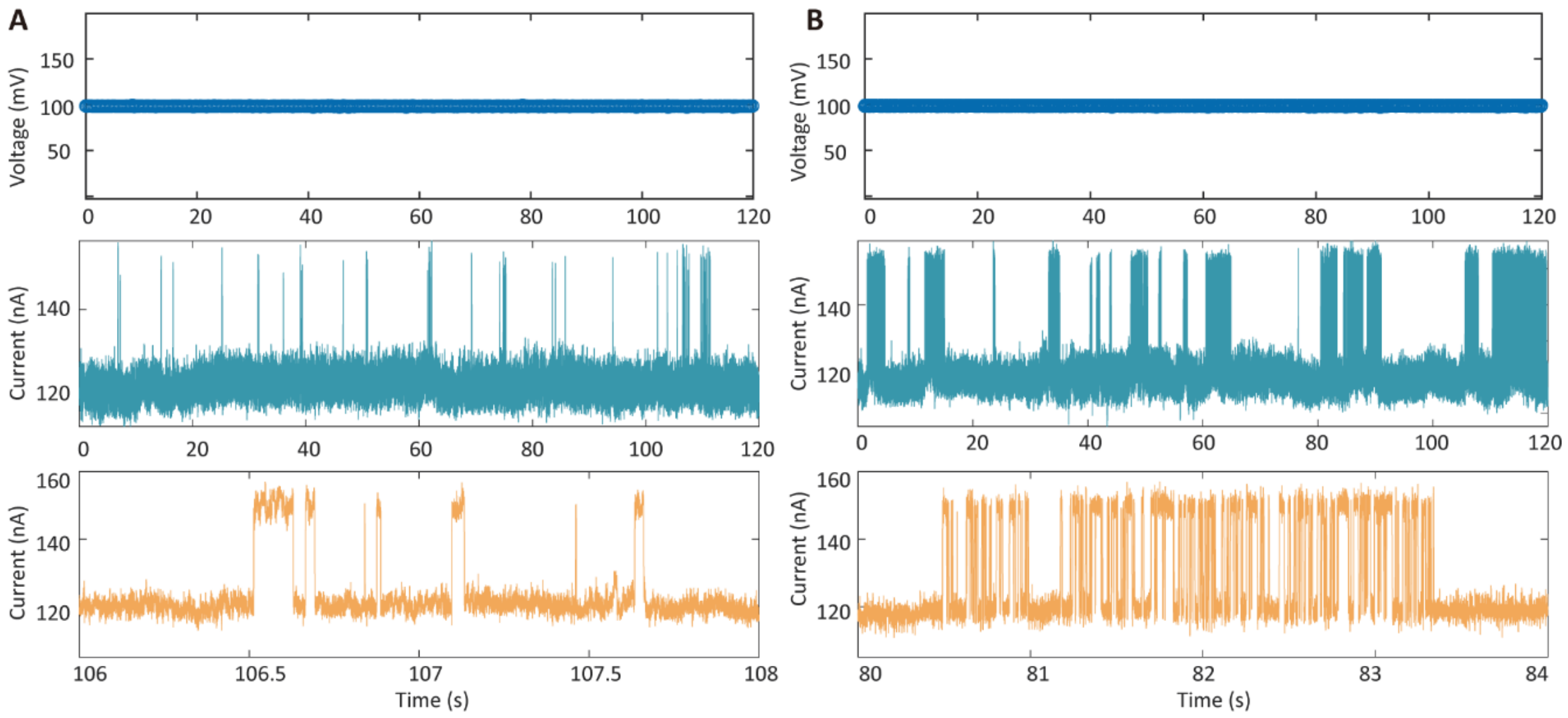

**Fig. S13.**

**Representative current-time traces of $[Ru(bpy)_3]^{2+}$ (A)** under dark conditions and **(B)** under light illumination.

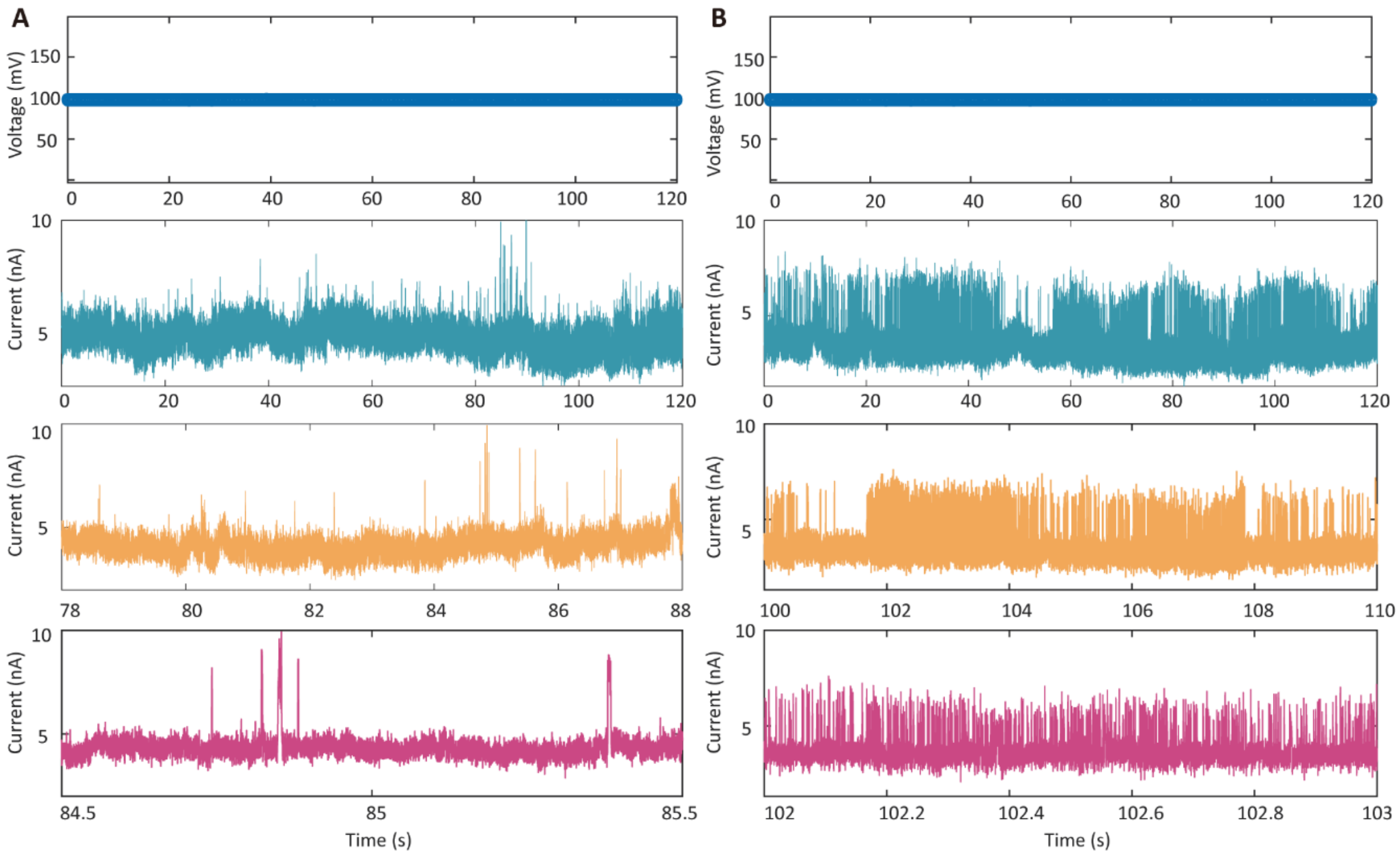

**Fig. S14.**

**Representative current-time traces of $C_{60}$ (A)** under dark conditions and **(B)** under light illumination.

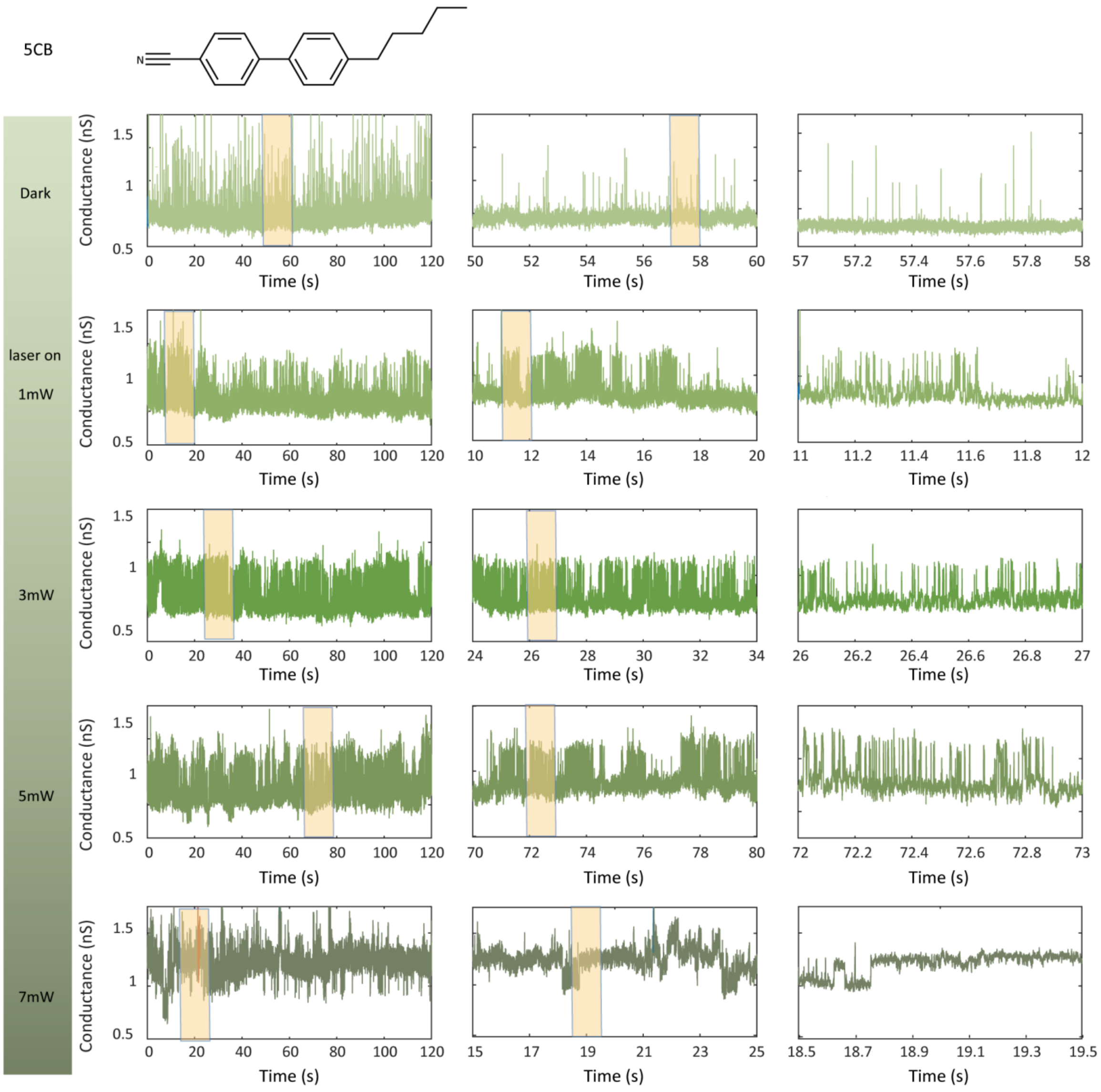

**Fig. S15.**

**Conductance traces of 5CB under varying laser power.** Representative conductance traces of 5CB acquired over different durations (120 s, 10 s, and 1 s) under varying laser powers. Increasing laser power led to a pronounced rise in the frequency of single-molecule conductance events, eventually resulting in a stabilised conductance state with extended dwell times. Furthermore, the conductance distribution narrowed with increasing laser power, attributed to enhanced local electromagnetic fields promoting a more ordered alignment of 5CB molecules within the tunnelling junction. Experimental conditions: 5CB concentration = 1 µM; applied bias = 100 mV; solvent = ethanol/water (1:1, v/v).

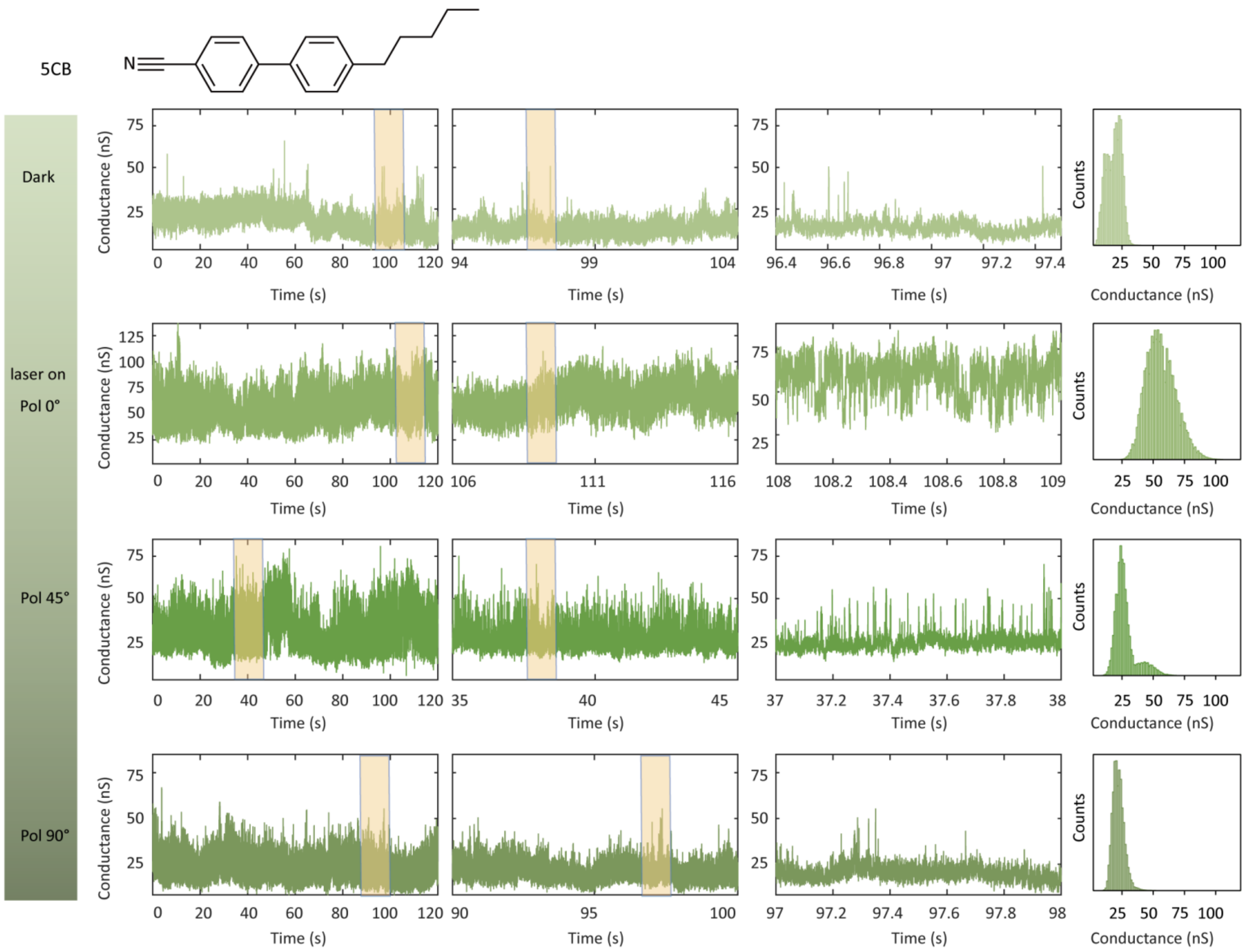

**Fig. S16.**

**Conductance traces of 5CB under varied polarisation conditions.** Representative conductance traces of 5CB recorded over durations of 120 s, 10 s, and 1 s under dark conditions and illumination with polarisation angles of 0°, 45°, and 90°. A polarisation angle of 0° corresponds to light polarised parallel to the coupling axis of the plasmonic tunnelling junction, yielding the greatest field enhancement for single-molecule plasmonic optical trapping. The QMT probes used in this polarisation-dependent experiment had larger gap distances compared to those used in the laser power-dependent measurements, resulting in higher baseline conductance values.

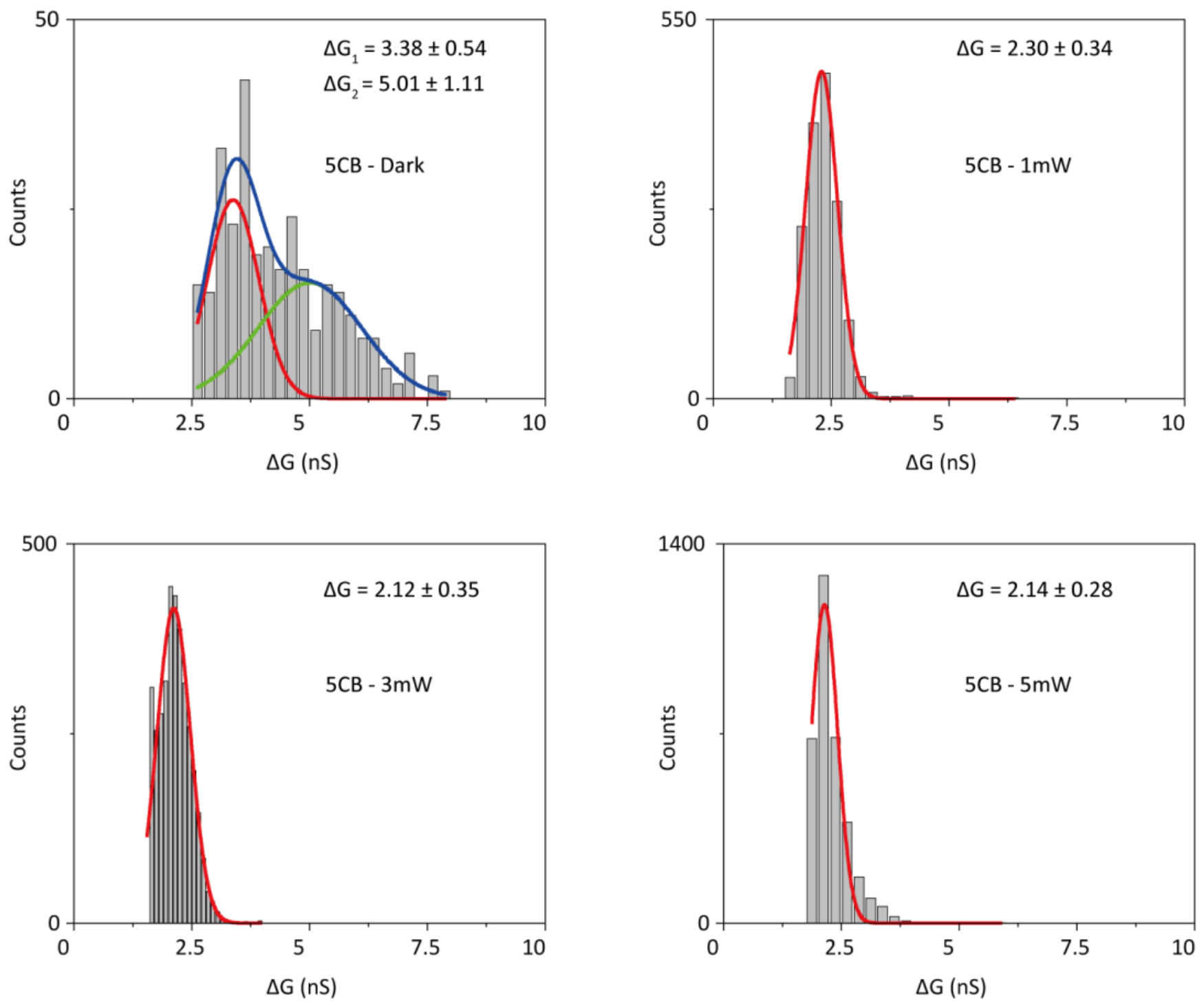

**Fig. S17.**

**The relative conductance (ΔG) distributions of 5CB under various laser powers.** The ΔG was determined by means of Gaussian fitting. The error bars represent ± SD from the means.

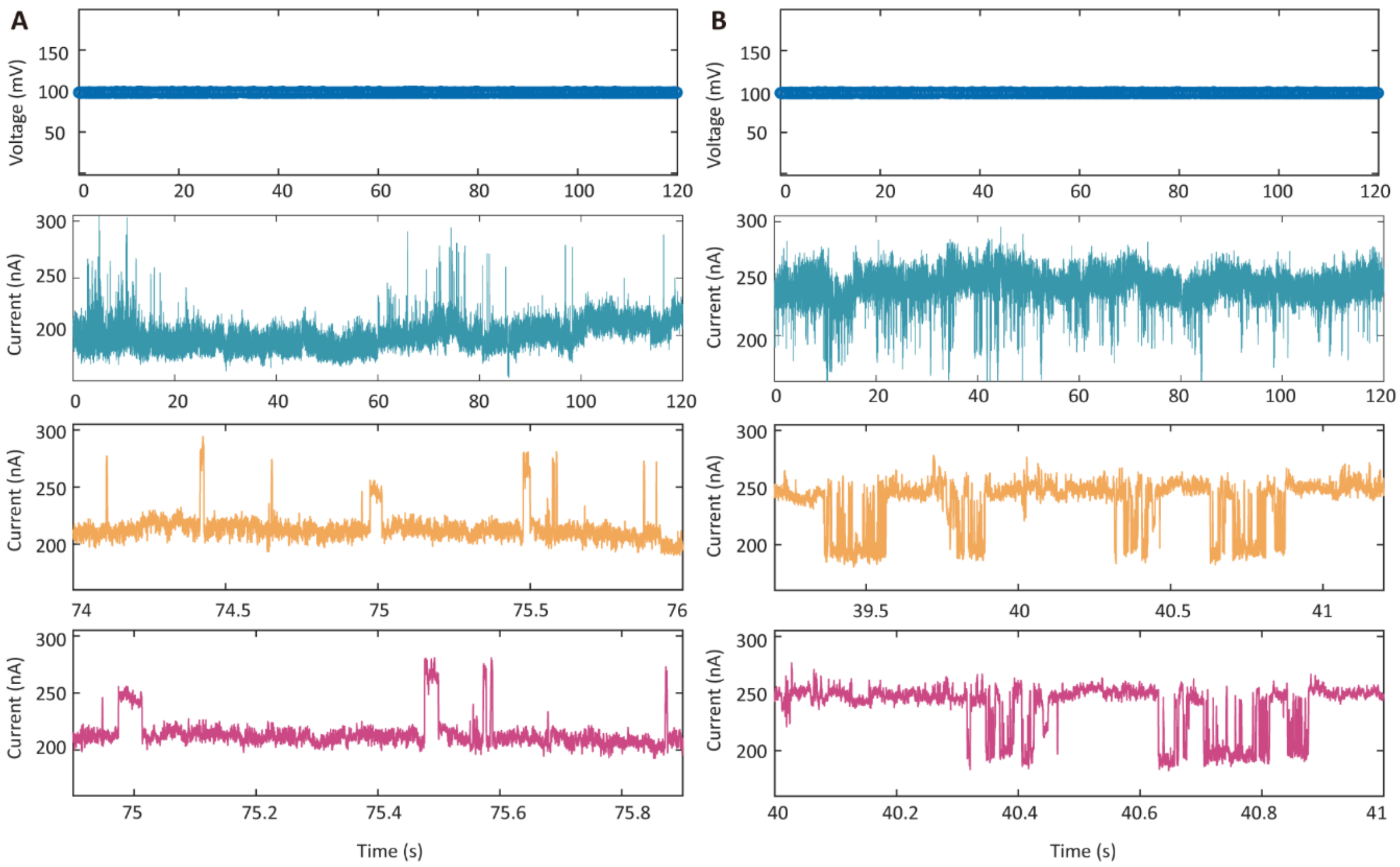

**Fig. S18.**

**Representative current-time traces of AuNPs (A)** under dark conditions and **(B)** under light illumination.

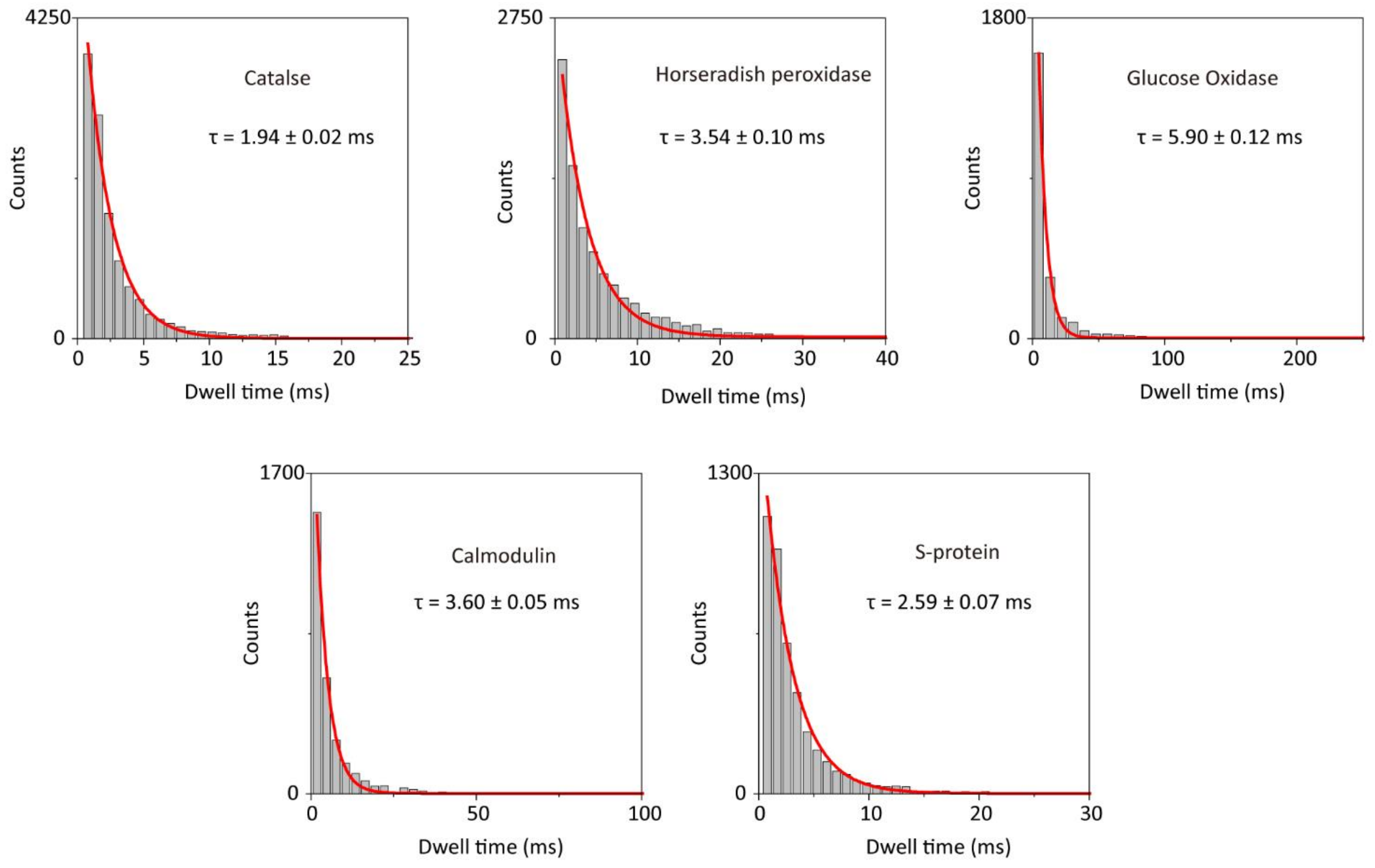

**Fig. S19.**

**Dwell time distribution of protein junctions.** Histograms of dwell times for individual protein junction events. The solid lines represent single-exponential fits to the data, indicative of Poissonian statistics consistent with stochastic single-molecule conductance events.

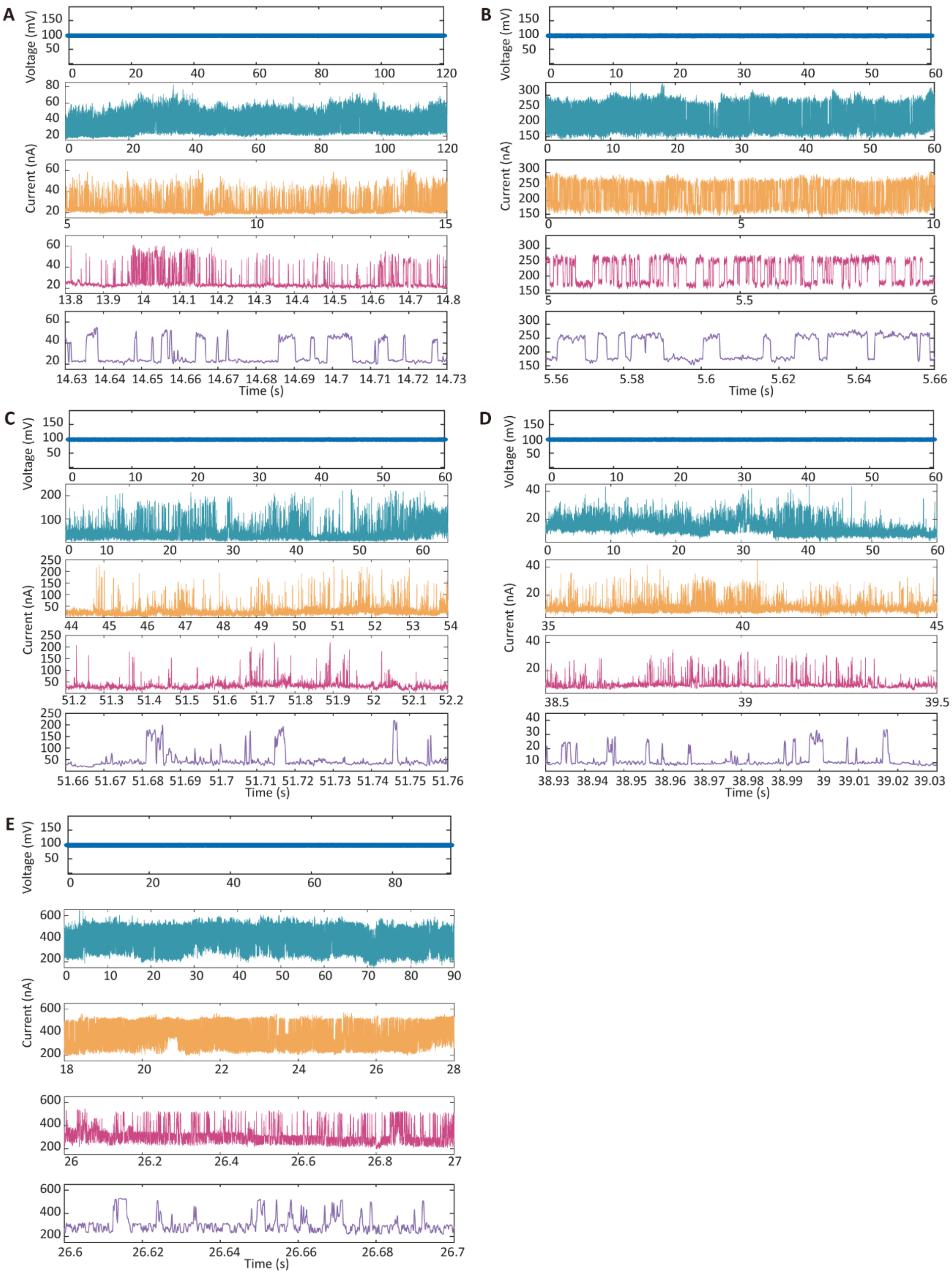

**Fig. S20.**

**Representative current-time traces of proteins under light illumination.** (**A**) catalase, (**B**) horseradish peroxidase, (**C**) glucose oxidase, (**D**) calmodulin, and (**E**) S-protein.

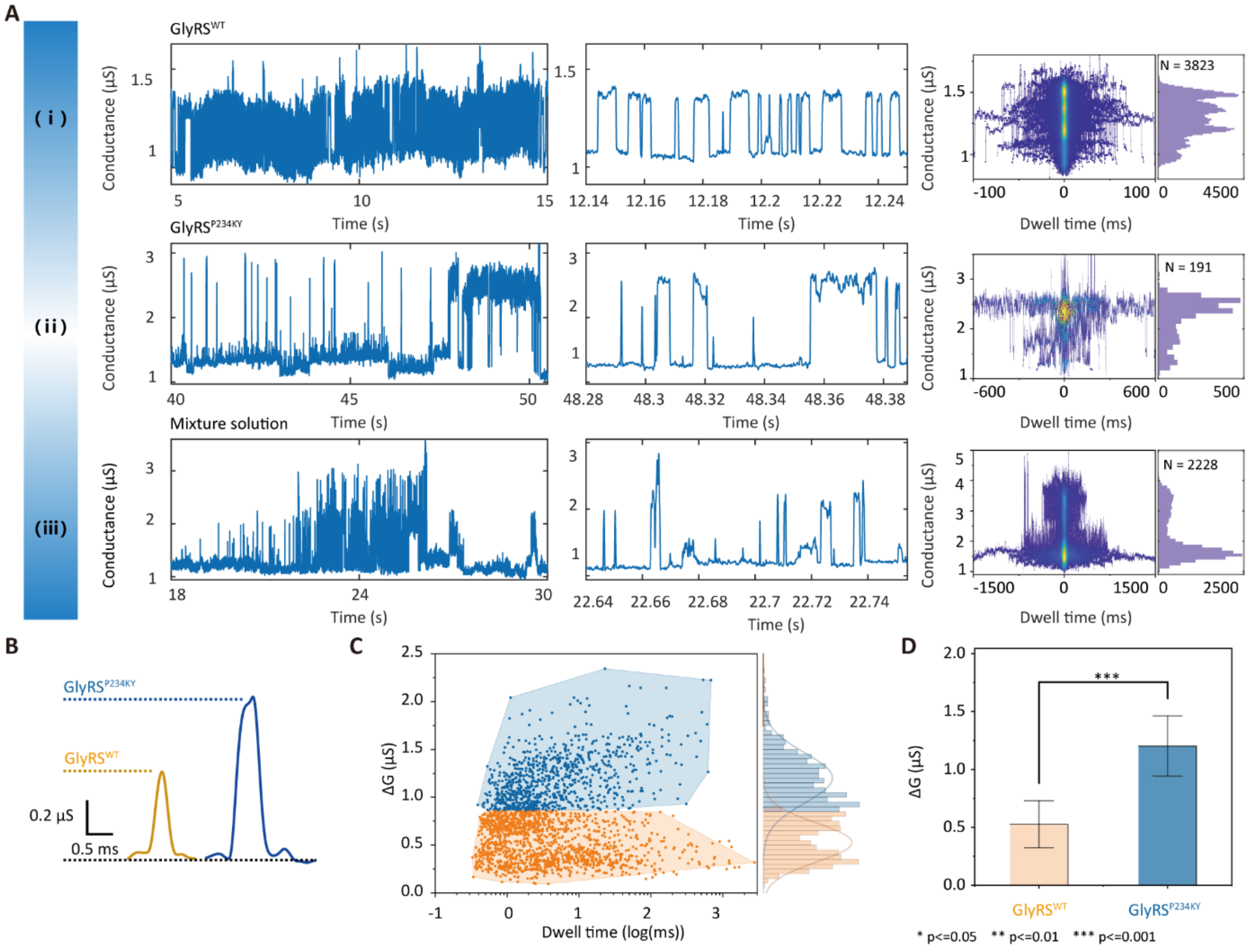

**Fig. S21.**

**Conductance profiling of GlyRS mutants (A)** Representative conductance traces and corresponding 2D density plots of peak signals for (i) GlyRS$^{WT}$, (ii) GlyRS$^{P234KY}$, and (iii) mixtures. Bias: 100 mV, laser power: 9 mW, buffer: 1 mM PBS, PH 7.4. **(B)** Typical peak signals of the low/high conductance states correspond to GlyRS$^{WT}$ (orange) and GlyRS$^{P234KY}$ (blue), respectively. **(C)** Scatter plot of ΔG versus dwell time (logarithmic scale) and accompanying histograms for a mixed protein sample of GlyRS$^{WT}$ and GlyRS$^{P234KY}$. The orange and blue dots denote the low and high conductance states extracted from the mixture measurement and classified via a clustering-based method. The histograms on the right panel show the distributions of two conductance states with Gaussian fitting. **(D)** Comparison of ΔG values for GlyRS$^{WT}$ (orange) and GlyRS$^{P234KY}$ (blue). Bars denote mean ΔG values; error bars indicate ±SD. Statistical significance was assessed via a two-tailed Student's t-test (**P < 0.001).

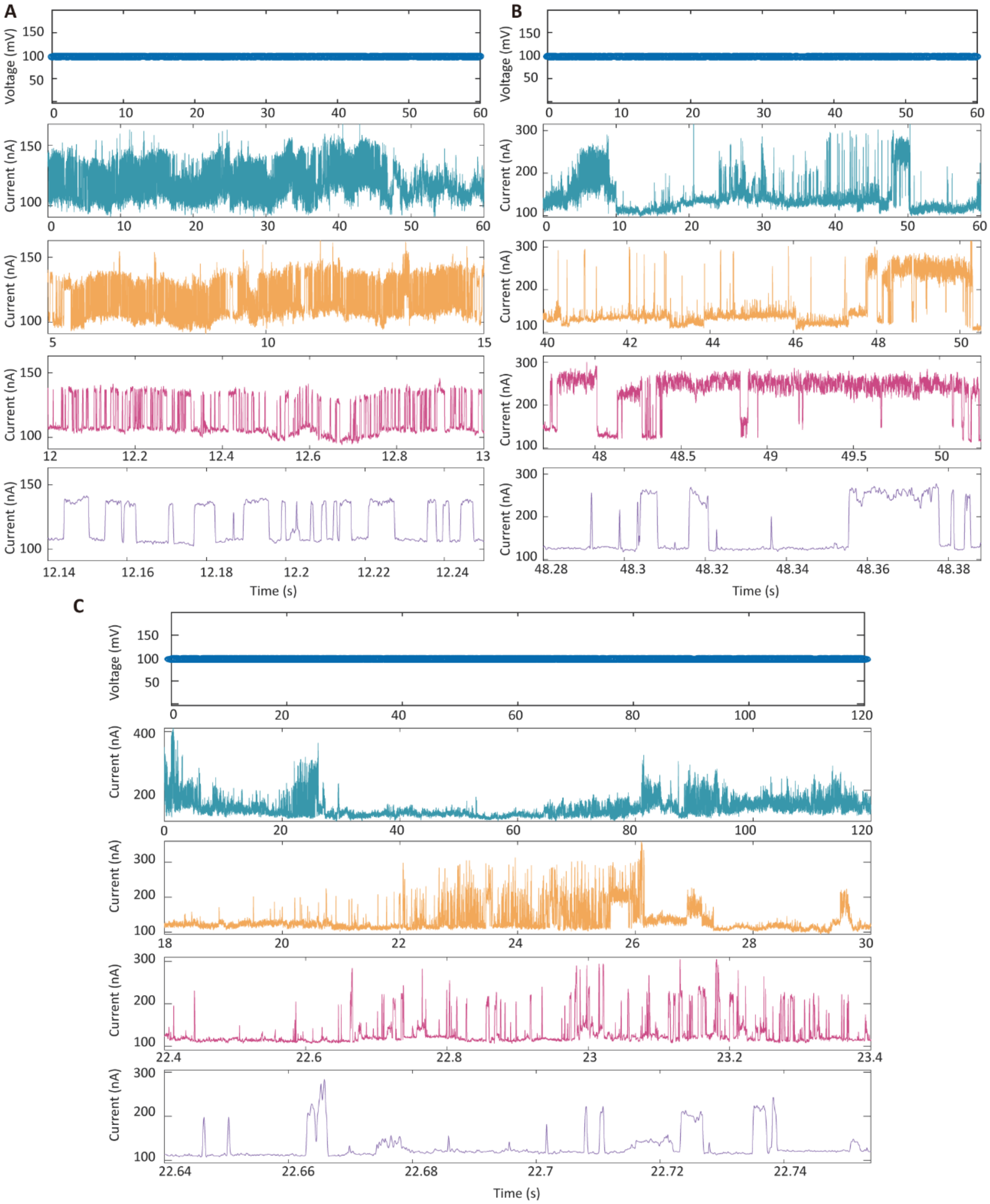

**Fig. S22.**

**Representative current-time traces of glycyl-tRNA synthetase under light illumination. (A)** GlyRS$^{WT}$, **(B)** GlyRS$^{P234KY}$, and **(C)** the mixture.

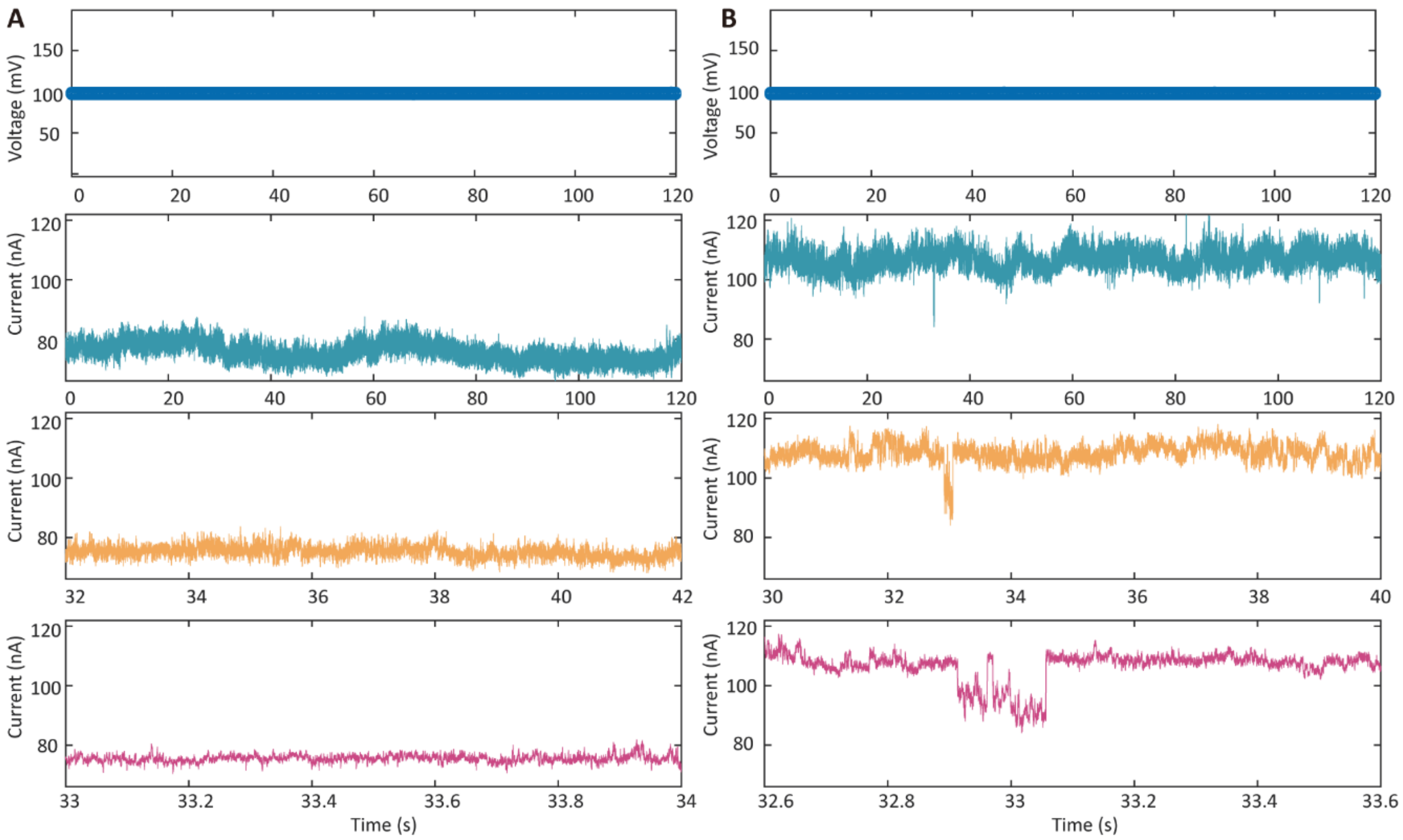

**Fig. S23.**

**Representative current-time traces of the His-tag-linked WDR5 protein junctions (A)** under dark conditions and **(B)** under light illumination.

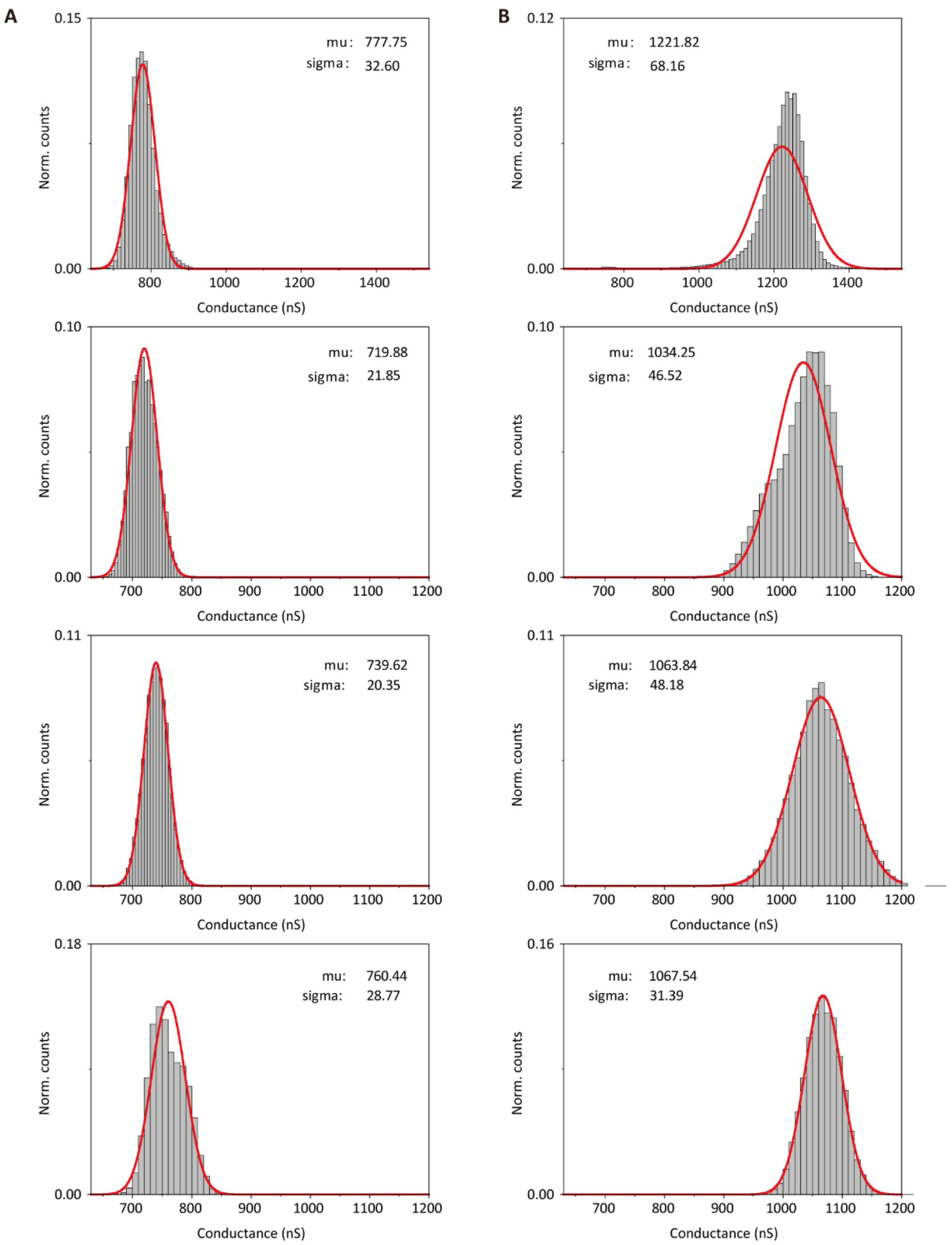

**Fig. S24.**

**The relative conductance (ΔG) distributions of the His-tag-linked WDR5 protein junction.** Histograms with Gaussian fitting show the ΔG of MSA-tag-linked WDR5 protein junction **(A)** with laser-off and **(B)** with laser-on. The inset shows ***mu*** and ***sigma*** obtained from Gaussian fitting, representing the mean and standard deviation of the ΔG distribution, respectively.

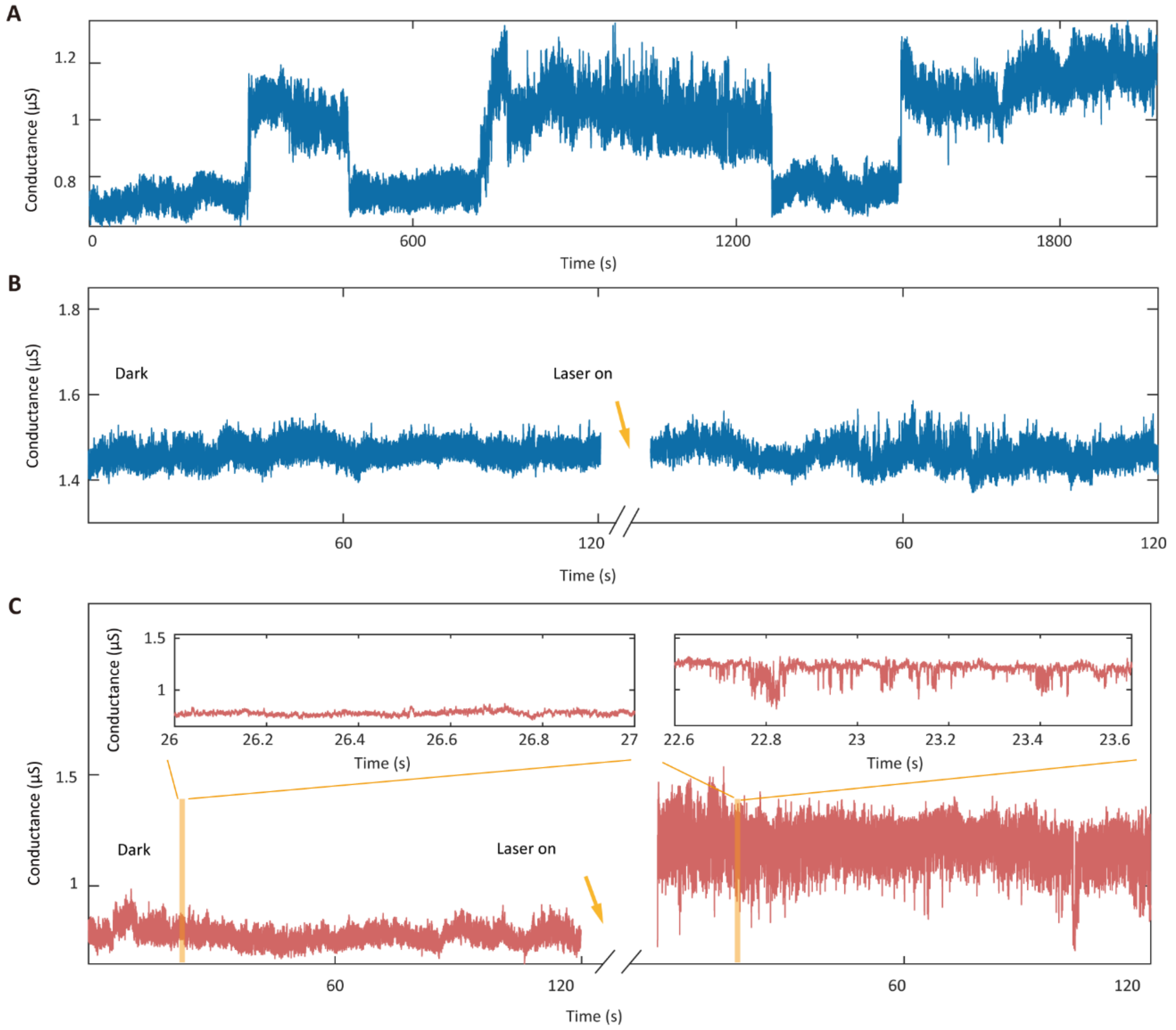

**Fig. S25.**

**Single-molecule switch of WDR5 protein junctions. (A)** Full conductance traces for trapping and releasing WDR5 protein junctions with His-tag corresponding to Fig. 4c. Bias: 100 mV, laser power: 5 mW. **(B)** Representative conductance traces of His-tag-linked WDR5 protein junctions after EDTA treatment, recorded under laser-off and laser-on conditions. The disappearance of signals indicates the disruption of protein junctions. Conditions: bias = 100 mV, laser power = 10 mW, buffer = 1 mM PBS, pH 7.4. **(C)** Representative conductance traces and zoomed-in views of 1-s traces for biotin-MSA-linked protein junction. Bias: 100 mV, laser power: 14 mW, buffer: 1 mM PBS, PH 7.4.

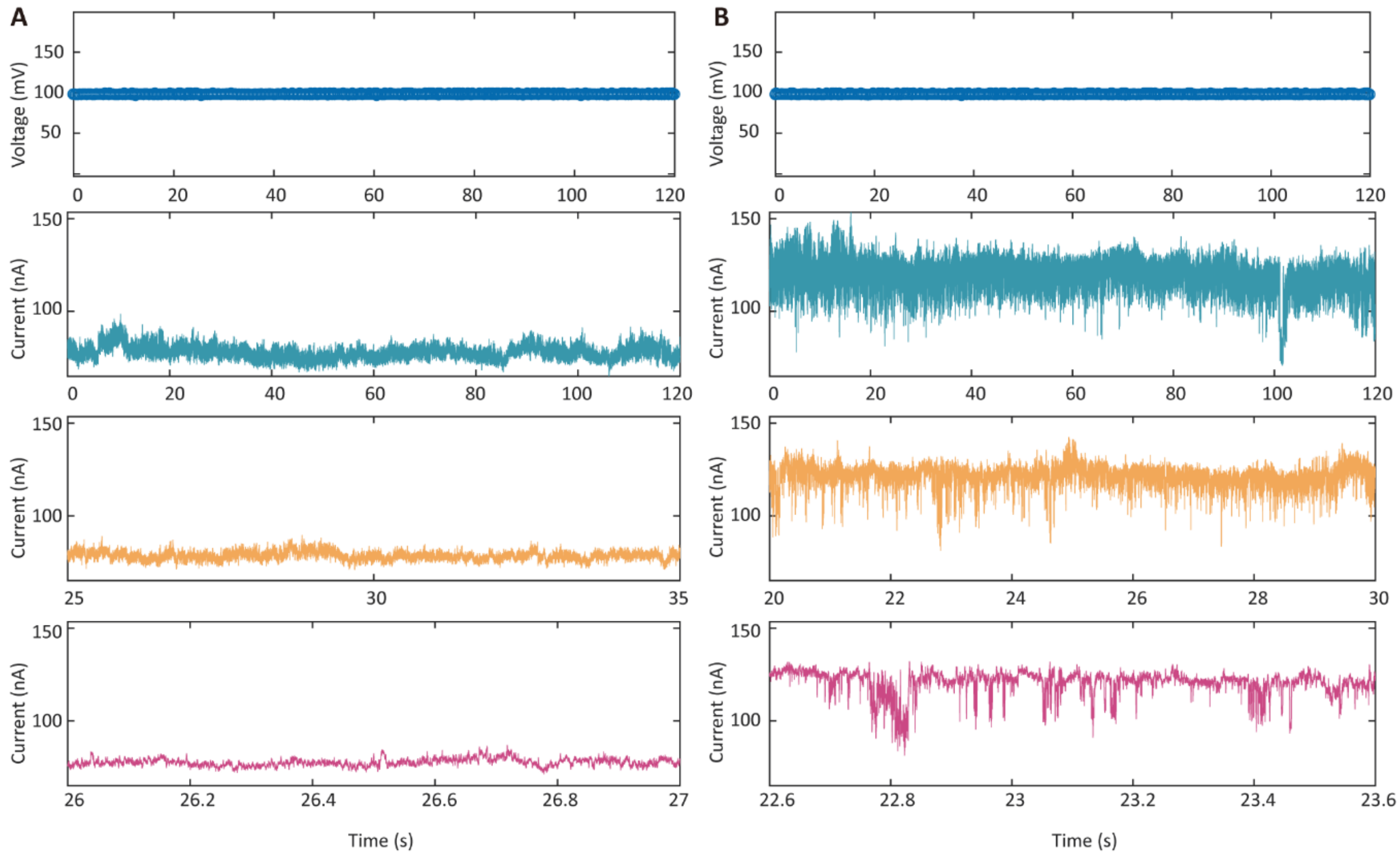

**Fig. S26.**

**Representative current-time traces of the MSA-tag-linked WDR5 protein junctions (A)** under dark conditions and **(B)** under light illumination.

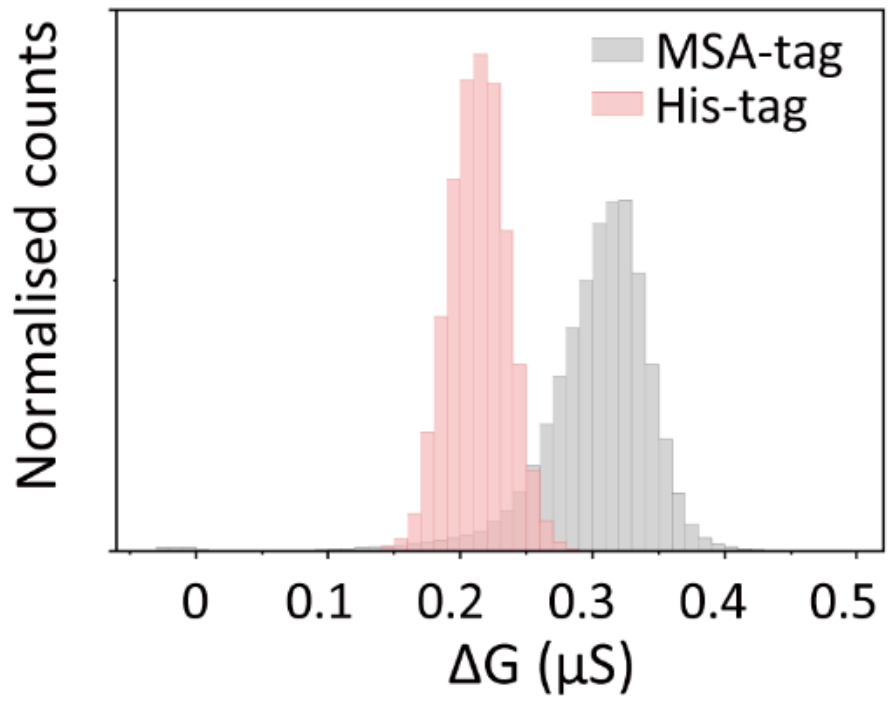

**Fig. S27.**

**Normalised histograms of ΔG for two types of WDR5-protein junctions**. Light blue and light red correspond to the Biotin-MSA-linked and His6-tag-linked WDR5 protein junctions, respectively.

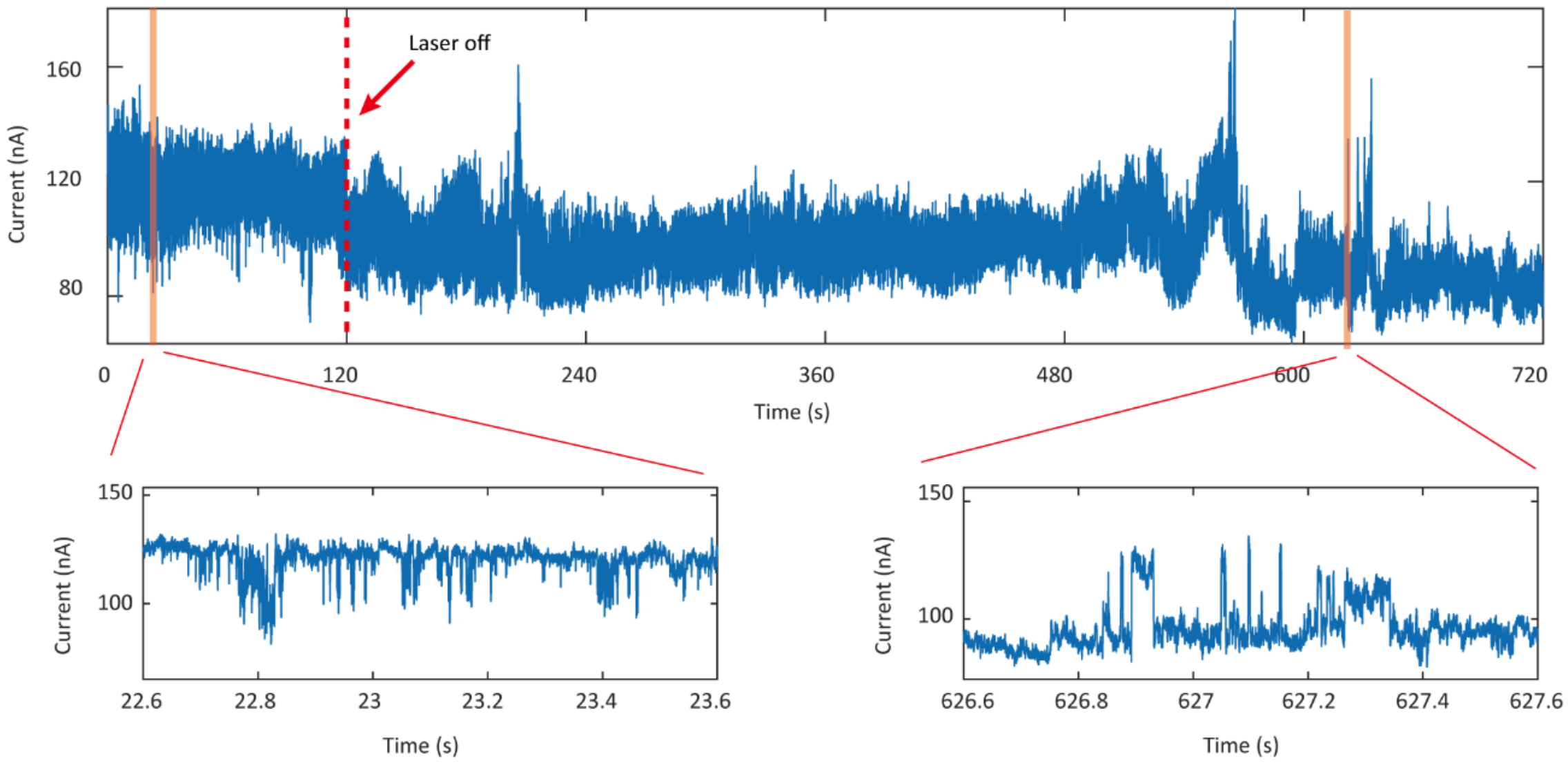

**Fig. S28.**

**Real-time current motoring of MSA-tag-linked WDR5 protein junction.** The red dashed line indicates the time point when the laser was turned off. The long-term continuous recording of current-time traces shows that the MSA-tag-linked WDR5 protein junction did not break immediately after the optical force disappeared. Bias: 100 mV, buffer: 1 mM PBS, PH 7.4.

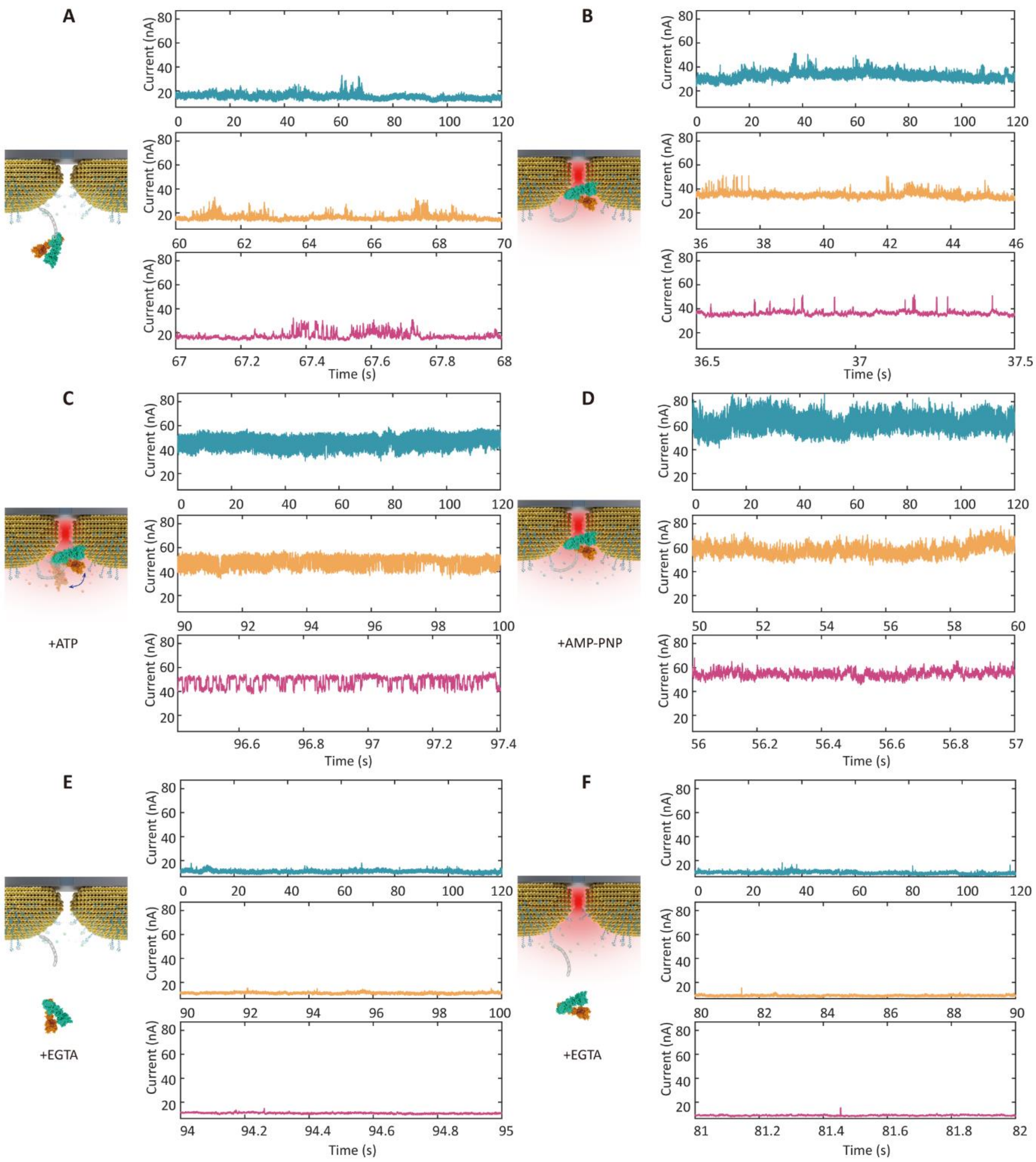

**Fig. S29.**

**Tracking protein conformational dynamics of Hsp90.** Representative current-time traces of Hsp90 protein junctions under various conditions: **(A)** laser-off, **(B)** laser-on, **(C)** ATP addition, **(D)** AMP-PNP (non-hydrolysable ATP analogue), and **(E-F)** chelation disruption by EDTA. Buffer: 1mM PBS, PH 7.4.

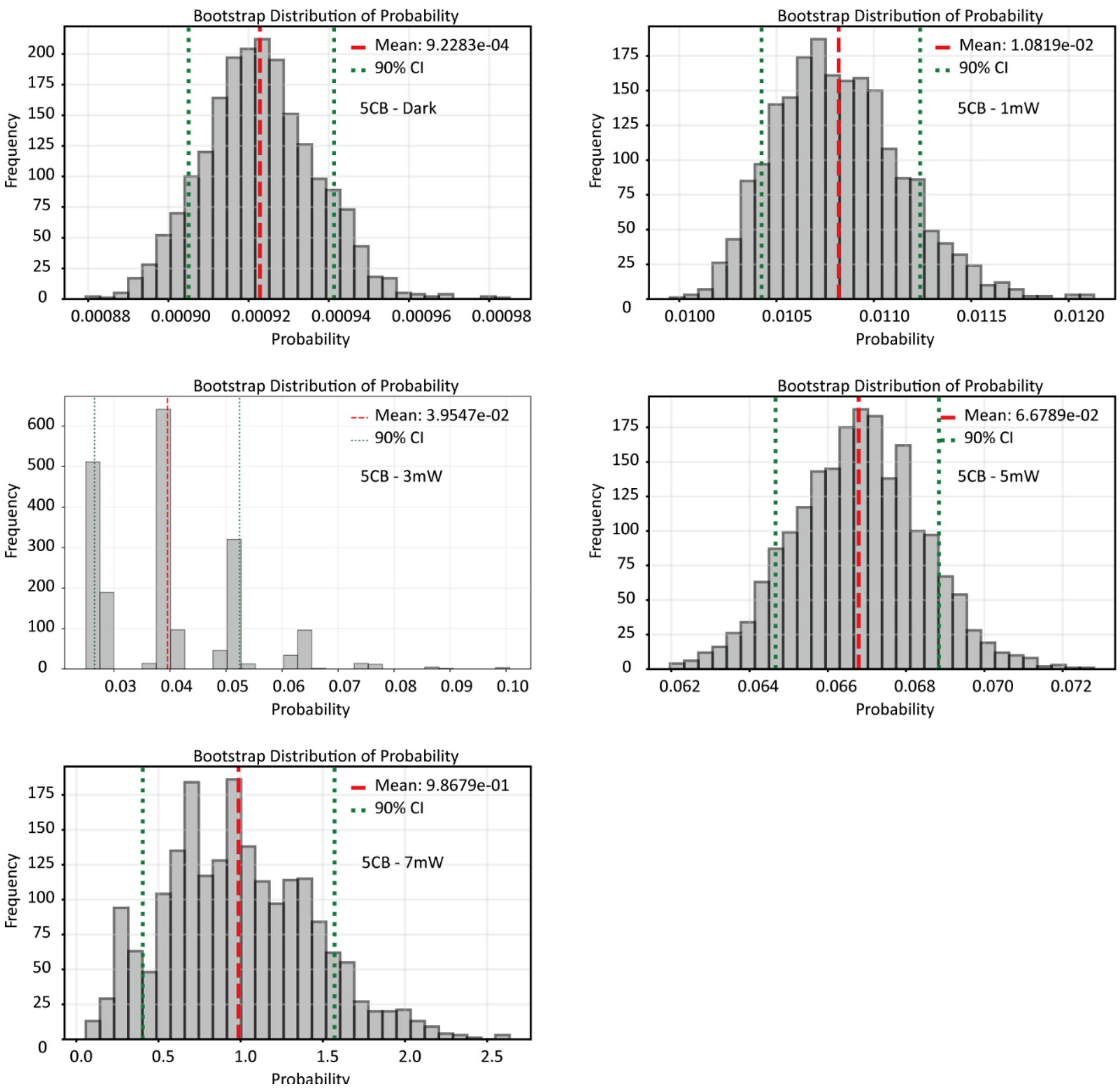

**Fig. S30.**

**Statistical analysis of the capture probability for 5CB under different light intensities.** Histograms show the bootstrapped capture probability distribution ($N_{boot} = 2000$) calculated as follows: first, extracting the dwell time of each signal peak that represents the single-molecule event. Then, the proportion of each single-molecule event in the total testing time (120-s traces) was calculated to form a large sample. Finally, the sum of all proportions, representing the capture probability, was statistically analysed through the bootstrapping method. The red and green dashed lines show the mean values and the 90% confidence interval of the bootstrapping.

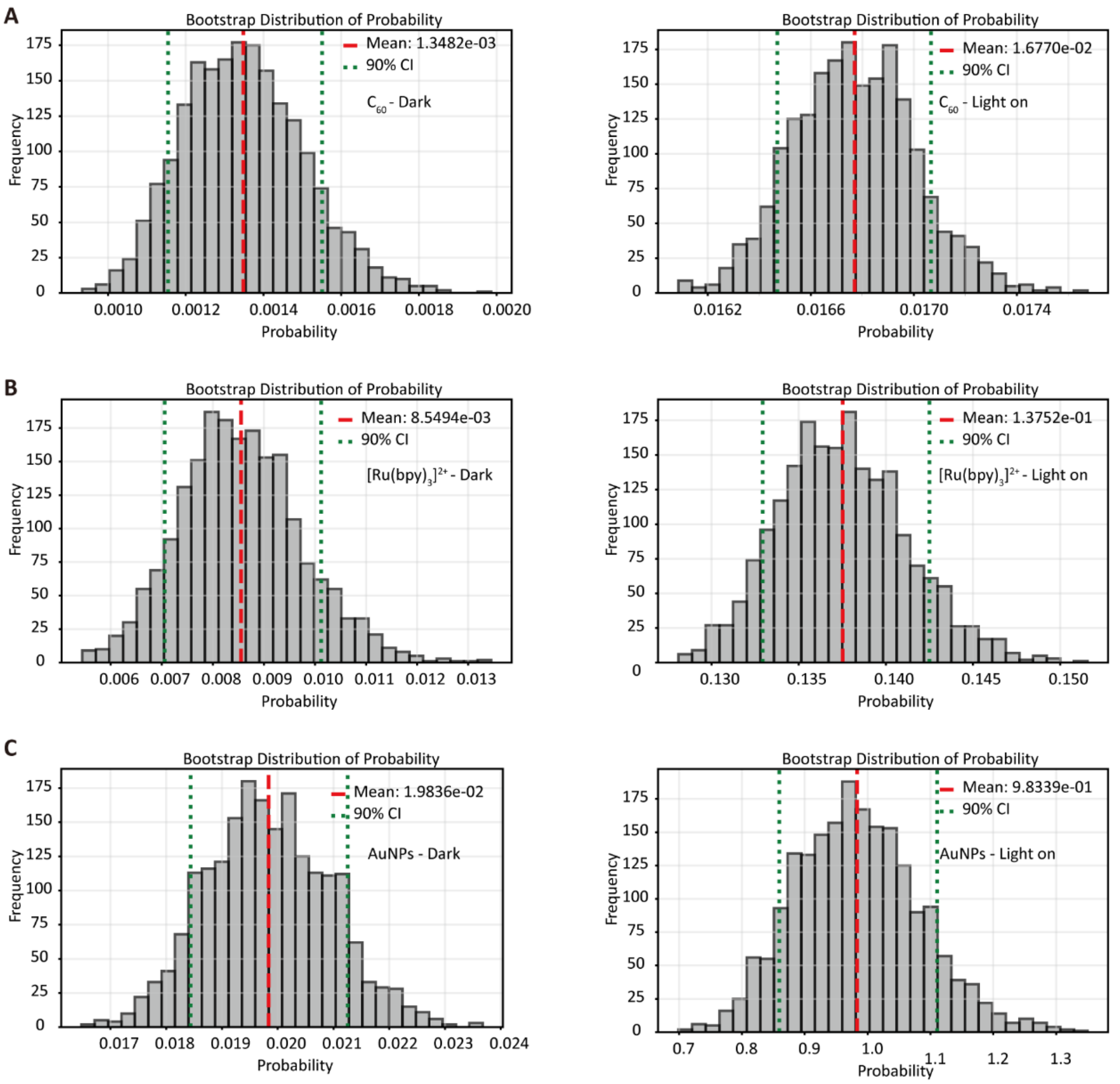

**Fig. S31.**

**Statistical analysis of the capture probability for $C_{60}$, $[Ru(bpy)_3]^{2+}$, and AuNPs.** Histograms show the bootstrapped capture probability distribution ($N_{boot} = 2000$) of **(a)** $C_{60}$, **(b)** $[Ru(bpy)_3]^{2+}$, and **(c)** AuNPs with and without laser illumination. The red and green dashed lines show the mean values and the 90% confidence interval of the bootstrapping.

**Table S1.**

**Summary of the protein junctions measured via the free collision strategy for conductance profiling.**

| Device | Protein | ΔG (μS) | τ (ms) | Gap distance (nm) | Barrier height (eV) |
|---|---|---|---|---|---|
| #1 | catalase | 0.2900. ± 0.0064 | 1.94 ± 0.02 | 1.29 | 0.59 |
| #2 | horseradish peroxidase | 0.8285 ± 0.1226 | 3.54 ± 0.10 | 0.70 | 2.33 |
| #3 | glucose oxidase | 0.5368 ± 0.0016<br>1.3273 ± 0.0299 | 5.90 ± 0.12 | 0.51 | 3.49 |
| #4 | calmodulin | 0.0638 ± 0.0280 | 3.60 ± 0.05 | 0.97 | 3.08 |
| #5 | S-protein | 1.5090 ± 0.3408 | 2.59 ± 0.07 | 0.63 | 3.24 |